# Complex Emotion Recognition System using basic emotions via Facial Expression, EEG, and ECG Signals: a review

Javad Hassannataj Joloudari [1,2,3], Mohammad Maftoun [4], Bahareh Nakisa [5], Roohallah Alizadehsani [6], Meisam Yadollahzadeh-Tabari [2]

[1] Department of Computer Engineering, Faculty of Engineering, University of Birjand, Iran

[2] Department of Computer Engineering, Babol Branch, Islamic Azad University, Babol, Iran

[3] Department of Computer Engineering, Technical and Vocational University (TVU), Tehran 4631964198, Iran

[4] Department of Artificial Intelligence, Technical and Engineering Faculty, South Tehran Branch, Islamic Azad University, Tehran, Iran

[5] School of Information Technology, Faculty of Science Engineering and Built Environment, Deakin University, Geelong, Vic, Australia

[6] Institute for Intelligent Systems Research and Innovation (IISRI) Deakin

University, Waurn Ponds, Australia

**Abstract:** The Complex Emotion Recognition System (CERS) deciphers complex emotional states by examining combinations of basic emotions expressed, their interconnections, and the dynamic variations. Through the utilization of advanced algorithms, CERS provides profound insights into emotional dynamics, facilitating a nuanced understanding and customized responses. The attainment of such a level of emotional recognition in machines necessitates the knowledge distillation and the comprehension of novel concepts akin to human cognition. The development of AI systems for discerning complex emotions poses a substantial challenge with significant implications for affective computing. Furthermore, obtaining a sizable dataset for CERS proves to be a daunting task due to the intricacies involved in capturing subtle emotions, necessitating specialized methods for data collection and processing. Incorporating physiological signals such as Electrocardiogram (ECG) and Electroencephalogram (EEG) can notably enhance CERS by furnishing valuable insights into the user's emotional state, enhancing the quality of datasets, and fortifying system dependability. A comprehensive literature review was conducted in this study to assess the efficacy of machine learning, deep learning, and meta-learning approaches in both basic and complex emotion recognition utilizing EEG, ECG signals, and facial expression datasets. The chosen research papers offer perspectives on potential applications, clinical implications, and results of CERSs, with the objective of promoting their acceptance and integration into clinical decision-making processes. This study highlights research gaps and challenges in understanding CERSs, encouraging further investigation by relevant studies and organizations. Lastly, the significance of meta-learning approaches in improving CERS performance and guiding future research endeavors is underscored.

## 1. Introduction

Affective computing is an interdisciplinary research domain that amalgamates the fundamental tenets of psychology, computer science, and cognitive science.
The automatic recognition of emotions is a crucial component of affective computing, whereby the recognition of emotional states serves as the foundation for the computer's comprehension of emotions and subsequent reactions [1]. The field of emotion recognition, which is highly appealing, has garnered significant attention from both industry and academia. Emotion recognition has numerous potential applications, including but not limited to human-computer interaction, video gaming, and continuous monitoring of infants and patients with medical conditions such as Parkinson's disease, Alzheimer's disease, depression, falls, and so on [2].
Basic emotions are primary, universal emotional states that are widely acknowledged in various societies and typically consist of happiness, sadness, fear, anger, surprise, and disgust [3]. These emotions are believed to be innate and have distinct facial expressions and physiological reactions [4]. In contrast, complex emotions are more intricate and frequently entail a blend of basic emotions alongside higher cognitive operations, such as jealousy, pride, and guilt [5]. Differing from basic emotions, complex emotions are shaped by personal encounters, social engagements, and societal standards [6].

In fact, emotion is a complex psychological phenomenon, characterized by distinct emotional states that manifest through various physical and physiological cues. These cues can be broadly categorized into physical expressions and physiological signals. Physical expressions include facial expressions, voice intonation, gait, and body posture. Physiological signals encompass indicators such as Electroencephalogram (EEG) and Electrocardiogram (ECG) readings, Galvanic Skin Response (GSR), and other biofeedback mechanisms. Additionally, emotions can be conveyed through written text and other communication methods, acting as mediators of our internal emotional states [1,7,8]. The recognition of emotions through physiological signals is an essential aspect of scrutinizing psychological states and advancing biofeedback-based applications. The advent of the metaverse concept and the integration of physiological signal trackers into smart devices have significantly advanced this area of investigation, making it both essential and intriguing for researchers.

Studies have shown that changes in emotional states can directly influence physiological signals, providing a valuable means of emotion detection.

The Autonomic Nervous System (ANS) and Central Nervous System (CNS) are essential in overseeing various physiological responses to emotional stimuli, like heart rate, skin conductance, and brain activity. Monitoring this regulation effectively can be done using techniques such as EEG and ECG. EEG allows for precise capture of dynamic changes in brain activity in response to emotional stimuli due to its high temporal resolution [9, 10]. In contrast, ECG provides a reliable measurement of heart rate variability, which is closely associated with emotional arousal and autonomic regulation [11, 12]). These methods together offer a comprehensive understanding of the physiological basis of emotional states, enhancing the analysis of facial expressions .

This understanding is vital for developing accurate and reliable methods for detecting and interpreting emotions through physiological signals [13, 14]. The objective of the researchers is to develop emotion recognition systems that exhibit equitable levels of accuracy and responsiveness.

The present difficulty in recognizing complex emotions stems from the limited array of physiological signals considered, which could obstruct a comprehensive understanding of the subtle emotional states that individuals undergo. Although physiological signals are crucial for emotion detection, the emphasis has frequently been confined to a restricted range of measurements that may fail to encapsulate the complexities of intricate emotions. Broadening the focus to include signals such as EEG, indicative of central nervous system activity, and ECG, which tracks heart activity influenced by the autonomic nervous system and sensitive to emotional intensity and stress, can offer a more comprehensive viewpoint. By merging these varied signals, it becomes feasible to reveal the detailed interactions between different physiological systems and complex emotional reactions, thereby improving both the precision and richness of emotion recognition while overcoming the constraints imposed by a limited signal emphasis [15].

However, given that these signals are non-linear and non-stationary, it is crucial to meticulously select appropriate features to significantly enhance the accuracy of the system [16, 17]. Another type of emotion recognition as the most effective, natural, and universal signal used by human beings to communicate their emotional states and objectives is facial expression [18]. From 1974, 55% of messages relating to emotions and beliefs are conveyed by facial expression, 7% through spoken words, and the remaining 40% are paralinguistic [19]. Numerous facial expression recognition (FER) approaches have been investigated in the fields of computer vision and machine learning to encode expression information from face representations.

Machine learning methodologies, such as Support Vector Machine (SVM), Multilayer Perceptron (MLP), Random Forest (RF), and K-Nearest Neighbor (KNN), are frequently utilized for the purpose of determining the presence of emotional content in facial expressions [20]. Nevertheless, the application of these methodologies extends beyond facial expressions alone, encompassing the analysis of physiological signals as well. This broader scope allows for a more comprehensive approach to the recognition of emotions. Through the integration of machine learning techniques with physiological signals (e.g., heart rate, skin conductance, brain activity) and physical cues (e.g., facial expressions), it becomes feasible to attain heightened precision and resilience in detecting emotions. This all-encompassing strategy capitalizes on the advantages offered by diverse data sources, ultimately facilitating a more profound comprehension of emotional states [21, 22 , 23].

Another variety of techniques that utilize deep architectures are known as deep learning methods, and they are outperforming traditional machine learning techniques in terms of accuracy and productivity. Deep Learning approaches, especially convolutional neural networks (CNN), which primarily depend on supervised learning using manually labeled data, have been inspected for facial emotion recognition [24]. A fixed learning algorithm that is manually created is usually used to train contemporary machine learning models from scratch for a specific task. Additionally, approaches based on deep learning have demonstrated remarkable success in numerous domains. There are, however, evident disadvantages. For instance, areas possessing massive computational power and the ability to accumulate and simulate massive quantities of data have seen the most success. Many applications where data is inherently expensive or rare or where computational resources are not available are disqualified by this [25 , 26].

In summary, the examination of basic emotions frequently involves the analysis of data gathered under controlled conditions, posing a notable obstacle. While such controlled settings enable precise evaluation of emotional reactions, they fall short in capturing the complex and varied nature of emotions encountered in real-life situations [27]. Real-world emotional responses are typically more complex and influenced by a myriad of factors, resulting in a fusion of basic and complex emotional states that defy easy classification. This disparity underscores a significant limitation: the datasets utilized for training deep learning models do not fully represent the emotional diversity encountered in daily life [28]. Moreover, deep learning models, despite their effectiveness, exhibit inherent constraints within this context. They often necessitate extensive sets of labeled data, which prove challenging to acquire for complex emotions given their subjective essence and the complexities involved in accurately labeling them [29]. This limitation hampers the models' capacity to generalize and excel in practical applications, where emotions are less predictable and more reliant on context. To tackle these hurdles, advanced deep learning methods and meta-learning strategies are under exploration. These approaches strive to enhance the flexibility and resilience of emotion recognition systems by utilizing smaller, more varied datasets and integrating contextually-aware learning mechanisms [30].

On the other hand, most famous datasets such as MultiPie, SFEW, RaFD, JAFFE, FER2013, etc are related to basic emotions while emotions are frequently more complicated than what basic emotions can communicate [31]. Thus complex emotions make keeping up with current emotion recognition systems difficult [32]. For complex emotion recognition systems due to the complexity and smallness of the data set, meta-learning approaches can play a vital role. The process of distilling the experience of multiple learning episodes often covering a distribution of related tasks and using this experience to improve future learning performance is called meta-learning [33, 34]. In recent years, few-shot learning [35], continual learning [36], label noises [37], and reinforcement learning [38] as the applications of meta-learning are commonly used in complex emotion recognition. The main goals of this paper are as follows:

This research delves into the enhancement of emotion detection through the amalgamation of facial expressions, EEG, and ECG signals using meta-learning techniques: The goal is to outperform conventional machine and deep learning approaches, thereby improving the adaptability and precision of emotion recognition systems for contextually aware real-world applications.

Pioneering the Study of Complex Emotion Recognition via AI: This paper marks the first in-depth investigation dedicated to understanding and developing AI systems specifically for complex emotion recognition, setting it apart from prior research that predominantly focused on basic emotions.

Examining the Contrasts Between Basic and Complex Emotions: We are delving into the fundamental discrepancies between basic emotions, which are universally acknowledged and simple, and complex emotions, which are intricate and influenced by the situation.

Providing Details About Complex Datasets: We offer a comprehensive overview of the datasets available for studying complex emotions, highlighting the challenges and intricacies involved in their use.

Assessing Existing Research on Basic Emotion Detection: By evaluating the body of research on machine learning and deep learning techniques used for basic emotion detection, we lay the groundwork for understanding the advancements and gaps that exist.

Categorizing Concepts for Complex Emotion Recognition through Meta-Learning: In this study, we categorize various approaches to complex emotion recognition based on meta-learning techniques, highlighting key methodologies that enhance the detection and understanding of emotions while expanding the capabilities of current AI technologies.

## 1.1 Publication analysis and search results

In this review paper, we examined 891 studies from various databases including IEEE, Science Direct, Springer, Wiley, and Google Scholar. This collection comprised 82 papers from IEEE, 60 from Springer, 158 from Science Direct, 139 from Wiley, and 450 from Google Scholar. In the initial phase, 801 records were eliminated due to duplication and irrelevance. Consequently, 90 records proceeded to the screening stage. During screening, 29 papers were excluded for being off-topic, leaving 61 publications for further review. At the eligibility stage, 23 more papers were removed. Ultimately, 38 research papers were selected as the final set of featured studies. The process of obtaining them based on PRISMA guidelines is shown in Fig. 1.

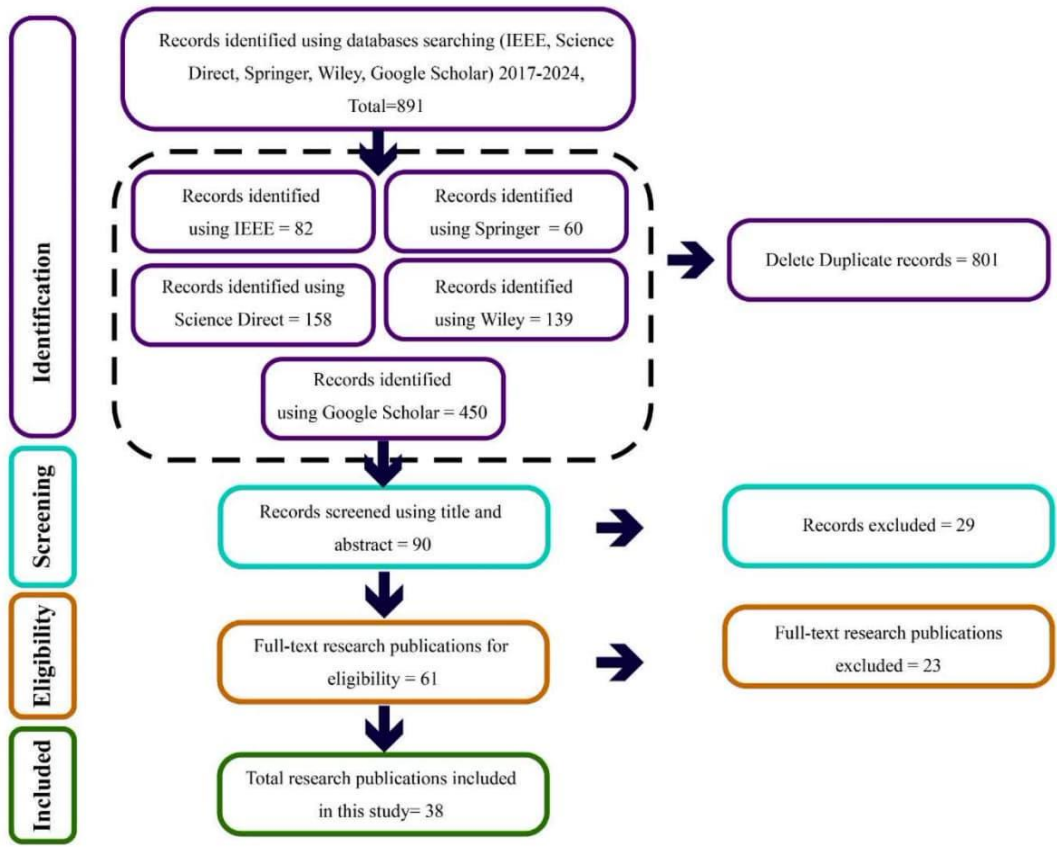

**Figure 1.** PRISMA flow diagram illustrating the selection process for relevant studies.

### 1.2 Data mapping of included studies

In this section, we provide the research papers selected from various databases. The trend chart for the number of publications on complex emotion recognition from 2017 to 2024 reveals a dynamic pattern as shown in Fig 2.

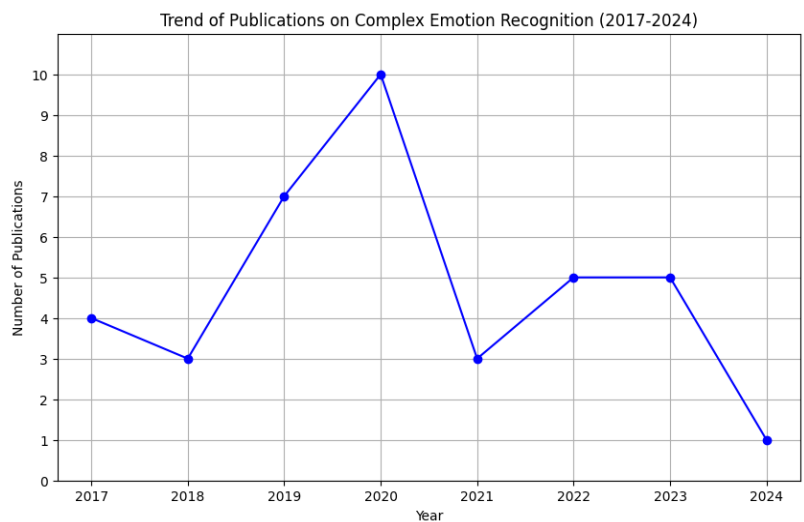

**Figure 2.** The trend of complex emotion recognition between 2017 and 2024.

From 2017 to 2020, there was a clear upward trend, with publications increasing from 4 to a peak of 10, indicating rising interest in the field. However, 2021 saw a sharp decline to 3 publications, possibly due to a temporary shift in research focus or resources. The numbers rebounded slightly in 2022 and 2023 with 5 publications each year, reflecting renewed efforts and interest. By 2024, the number dropped to its lowest at 1 publication, suggesting a potential winding down of major projects or shifting research priorities. Overall, the data highlights fluctuating engagement in complex emotion recognition research over the years.

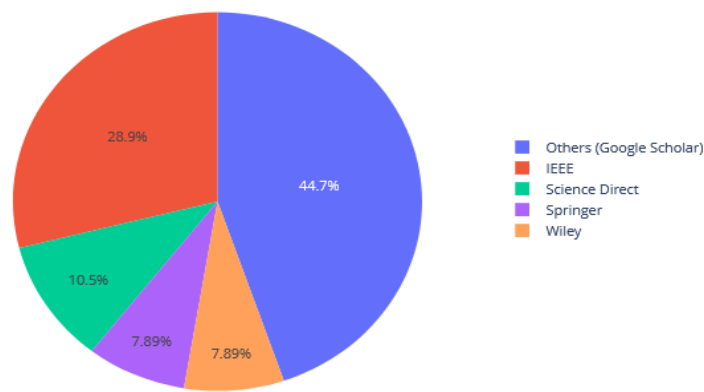

**Figure 3.** The allocated portion of each database in the complex emotion recognition topic.

According to Fig 3, the Google Scholar (others) database houses the largest proportion of research papers in this field, accounting for nearly half (44.7%) of the total publications. The IEEE database follows, with 30.9% of the publications. Additionally, 10.5% of relevant studies were found in Science Direct. Lastly, while the Wiley and Springer databases each contributed only 7.89% of the research papers, they have nonetheless published valuable studies on complex emotion recognition systems.

The next step of this study is structured as follows: Section 2 describes the type of emotions, the differences between them and emotion models In section 3, the complex emotion databases are reviewed and analyzed. Section 4 the preprocessing steps of developing a complex emotion recognition system are presented. Section 5 focuses on the concepts and technologies of emotion recognition systems and also limitations of traditional methods. In section 6, a review of related works are conducted. Section 7, the discussion and evaluation metrics are delivered.In Section 8, and last section are addressed into Open Research Challenges , Conclusions and future research directions.

## 2. Type of emotions, differences and emotion models

### 2.1 Basic Emotions

Emotions that are basic are brief emotional reactions that often occur automatically and have counterparts in other vertebrates. They are generally recognized worldwide and are commonly linked to specific physical reactions and facial expressions. Fear, anger, disgust, sadness, joy, and surprise are among the most frequently acknowledged basic emotions [39]. Key features of basic emotions comprise:

1. **Automaticity**: Basic emotions are activated automatically in reaction to particular stimuli without conscious deliberation. For instance, the sight of a predator may trigger fear, while the presence of a loved one may trigger joy.
2. **Distinctive Facial Expressions**: Each basic emotion is tied to a distinct facial expression that is easily recognized across diverse cultures. For instance, a frown and squinting eyes are typically associated with anger, while a grin signifies happiness. Alongside facial signals, physiological metrics like EEG and ECG offer a more profound insight into emotional experiences. The EEG method captures the brain's activities tied to various emotional states, while the ECG technique evaluates heart rate variability and autonomic nervous system responses, both being notably affected by emotional excitement and stress. Collectively, these indicators present a more comprehensive view of emotion, merging external signals with internal physiological reactions.

3. **Physiological Responses**: Basic emotions come with specific physiological alterations in the body, such as variations in heart rate, breathing, and hormone levels. Fear, for example, initiates the "fight or flight" response, resulting in heightened heart rate and adrenaline release.
4. **Evolutionary Roots**: Basic emotions are thought to have distinct evolutionary functions tied to survival and procreation. Fear, for instance, aids individuals in reacting to dangers, while joy strengthens social connections and motivates actions that enhance overall well-being.

**2.2 Complex Emotions**

Complex emotions are intricate and multifaceted emotional experiences that encompass a blend of basic emotions, cognitive processes, and social influences. Unlike basic emotions, which are generally widespread, complex emotions can vary significantly among individuals and societies [40]. Here are the primary characteristics:

1. **Cognitive Components:** Complex emotions entail substantial cognitive assessment and understanding of situations. Individuals may have to evaluate the context, their personal beliefs and values, and the perspectives of others to fully understand and experience intricate feelings. For instance, feelings of jealousy may emerge from interpreting a situation as a threat to a valued relationship.
2. **Cultural and Individual Variation:** In contrast to basic emotions, which are predominantly uniform, complex emotions can differ greatly among individuals and cultures. Cultural norms, values, and personal experiences have a substantial impact on shaping the experience and expression of complex emotions. For example, the way pride is felt and exhibited may vary across cultures.
3. **Extended Duration:** Complex emotions typically endure longer than basic emotions and may transform over time as new information and experiences are assimilated. For example, feelings of guilt may persist as individuals contemplate their actions and their repercussions.
4. **Interpersonal Functions:** Complex emotions often serve a vital role in managing social connections and navigating social standards. They assist individuals in comprehending and responding to others' emotions, negotiating social hierarchies, and upholding social ties. For instance, expressions of romantic love may entail a complex interplay of emotions like affection, desire, and commitment.

Basic emotions and complex emotions differ in several key aspects. Firstly, regarding duration, basic emotions are typically short-lived and immediate, whereas complex emotions tend to persist for extended periods. Secondly, in terms of cognitive involvement, basic emotions often arise automatically with minimal cognitive processing, while complex emotions require significant cognitive appraisal and interpretation of situations. Thirdly, basic emotions have clear evolutionary functions primarily related to survival, whereas complex emotions have evolved to manage more sophisticated social interactions and relationships. Finally, basic emotions are universally recognized and expressed similarly across cultures, whereas complex emotions can vary significantly based on cultural and individual differences in expression and recognition [41,42,43].

**2.3 Differences Between Basic and Complex Emotions**

To further clarify the differences between basic and complex emotions, it is crucial to explore their cognitive and social implications more deeply . Basic emotions like joy, fear, anger, sadness, disgust, and surprise are often seen as universal and innate, arising automatically in response to stimuli with little conscious thought. These emotions are believed to have developed to fulfill specific survival purposes, such as fear activating fight-or-flight reactions to possible dangers, and disgust aiding individuals in avoiding harmful substances. Studies by [3] and others have shown that basic emotions are expressed and understood similarly across various cultures, hinting at a common evolutionary origin.

In contrast, complex emotions such as guilt, shame, envy, pride, and love involve more intricate cognitive processes, often requiring self-reflection, comparison with others, and an understanding of societal norms and expectations. These emotions are not only enduring but also dependent on context, influenced by personal encounters, cultural upbringing, and the particular social setting. For example, feeling shame involves evaluating one's actions in relation to societal standards and anticipating others' judgments, which can differ significantly among cultures [44, 45]. Research indicates that while the fundamental elements of these emotions might be universal, their expression, interpretation, and importance are molded by cultural context and individual distinctions [46].

Furthermore, complex emotions frequently result from combinations or interactions of basic emotions [47] and are more likely to involve mixed sentiments or uncertainty. For instance, jealousy could blend fear (of losing a relationship) with anger (towards a perceived rival) and sadness (due to feeling unappreciated). This complex interplay of emotions showcases the sophisticated cognitive processes that underlie complex emotions, making them more challenging to investigate and comprehend compared to basic emotions [48]. Additionally, the influence of language in expressing and shaping complex emotions should not be underestimated; different societies possess specific terms and ideas that capture nuanced emotional states, emphasizing further the diversity in how these emotions are felt and conveyed [49].

Basic emotions and complex emotions differ in several key aspects. Firstly, regarding duration, basic emotions are typically short-lived and immediate, whereas complex emotions tend to persist for extended periods. Secondly, in terms of cognitive involvement, basic emotions often arise automatically with minimal cognitive processing, while complex emotions require significant cognitive appraisal and interpretation of situations. Thirdly, basic emotions have clear evolutionary functions primarily related to survival, whereas complex emotions have evolved to manage more sophisticated social interactions and relationships. Finally, basic emotions are universally recognized and expressed similarly across cultures, whereas complex emotions can vary significantly based on cultural and individual differences in expression and recognition [50,51,52,53,54].

### 2.4 Emotion models

Defining emotion, or affect, is essential in setting criteria in affective computing. The understanding of emotions was initially presented by Ekman in the 1970s. Despite psychologists' various attempts to classify emotions in fields like neuroscience, philosophy, and computer science, there is no universally agreed-upon model of emotions. However, the two main types of emotion models commonly utilized are the discrete (or categorical) emotion model and the dimensional (or continuous) emotion model [55].

In the discrete emotion modelas which illustrated by the wheel of emotions depicted in Fig 4, individuals typically choose one emotion from a set of predefined emotions that best represents their feelings. This model categorizes emotions clearly and is based on Ekman's early research.

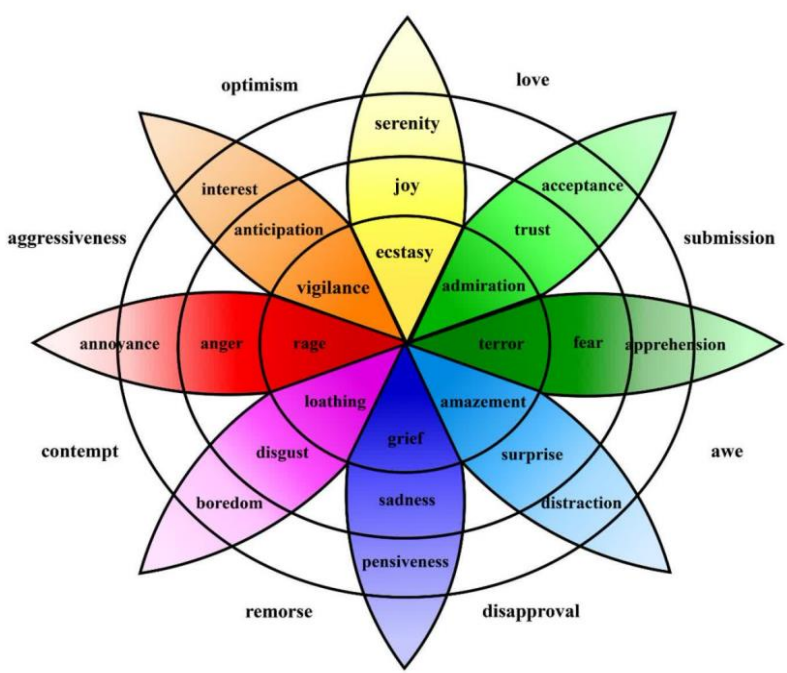

**Figure 4.** Wheel of emotions based on discrete models.

Conversely as shown in Fig 5, the dimensional model portrays emotions using quantitative measures across various dimensions. This method often utilizes tools such as the Self-Assessment Manikin (SAM) or Feeltrace. SAM uses images of SAMmanikins to evaluate the static level of a dimension at a specific moment, while Feeltrace monitors emotional data continuously over time [56].

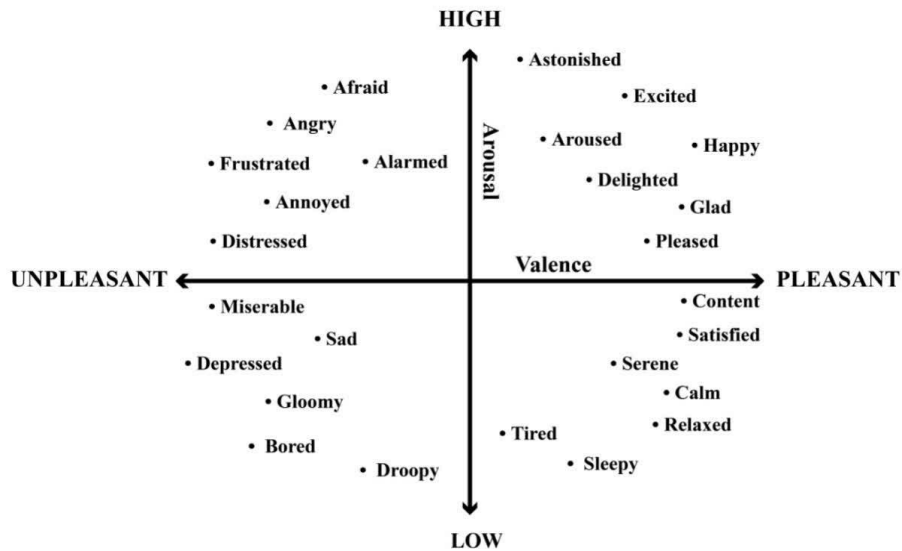

**Figure 5.** Emotions based on dimensional models.

Both models emphasize different facets of human emotion and provide insights into how emotions are perceived and understood by the human mind. The discrete model sorts emotions into categories with distinct labels, while the dimensional model captures the intricacy and variability of emotional states through continuous assessment. Together, these models contribute to a thorough comprehension of genuine emotional states and are crucial in the realm of affective computing.

### 3. Datasets used in scientific studies

Datasets are fundamental for developing every AI-based system, encompassing the essential attributes necessary for facial emotion recognition, as well as for the analysis of EEG and ECG signals. These datasets facilitate the extraction and interpretation of results, and they are combined in advance to support the development of algorithms for recognizing facial emotions and analyzing physiological signals. They include a variety of data types such as static images, videos, or a combination of both [57 , 58]. Additionally, some datasets include pictures taken in staged settings using laboratory-controlled environments or are provided by psychology centers. Moreover, datasets derived organically from real-world environments are also available.

#### 3.1 Basic and complex Emotion Datasets

**CMED**

Based on existing spontaneous micro-expression datasets such as CASME I, CASME II, CAS(ME), and SAMM, researchers created the Compound Micro-expression Database (CMED). These datasets mainly gather facial expression data rather than physiological signals. Due to their subtle features, it is challenging to analyze the motion track and characteristics of micro-expressions. Consequently, generating compound micro-expression images presents numerous obstacles and limitations. The data in these datasets is captured using high-speed cameras that record slight facial movements at a high frame rate, typically ranging from 100 to 200 frames per second, enabling the detection of micro-expressions lasting between 1/25th and 1/5th of a second. Participants in these datasets engage in various tasks or are exposed to specific stimuli to evoke spontaneous emotional reactions, which are then documented [59 ,60, 61, 62, 63]. The number of participants varies among the datasets. For example, the CASME II dataset contains recordings from 26 individuals, while the SAMM dataset includes data from 32 subjects. The exact number of participants in the CMED may differ, but it usually encompasses a diverse group to ensure a wide array of facial expressions and emotions. These datasets cover a range of both basic and complex emotions. Common basic emotions captured include happiness, sadness, anger, fear, surprise, and disgust, which are universally understood and easily recognizable. Additionally, complex emotions like contempt, embarrassment, pride, and guilt are also recorded to offer a more thorough insight into human emotional expression.

**CASME**

The first dataset established by the Chinese Academy of Sciences (CAS) was CASME. It was recorded at 60 frames per second and includes 180 videos from 19 subjects. Thirteen females and twenty-two males, with an average age of 22.03 years ($\sigma = 1.60$) were enlisted for the study. The eight micro-expressions that have been recognized are: amusement, sadness, disgust, surprise, contempt, fear, repression, and tense. The experiments were limited to the top five classes (tense, disgust, happiness, surprise, and repression) with the greatest number of samples[64].

**CFEE**

This dataset contains 5,060 images of faces from 230 people which have been labeled with 15 complex emotions and 7 basic emotions.[65] There are 21 different emotion categories in the CFEE dataset. Compound emotions, like delightfully surprised and angry surprised, comprise multiple basic component categories. The Facial Action Coding System was utilized for the analysis of the obtained images. These 21 categories have distinct production processes that are nevertheless in line with the subordinate categories they represent[66].

**FER-2013**

This dataset was introduced at the 2013 International Conference on Machine Learning (ICML) by Pierre-Luc Carrier and Aaron Courvill, the 2013 Facial Expression Recognition dataset (FER-2013) is available in the Kaggle dataset. The 35,887 grayscale 48x48-pixel images in the FER-2013 dataset remain in a spreadsheet with the pixel values of each image listed in a row of cells. After using Google to source images, they were categorized into various emotional classes including surprise, anger, disgust, fear, happiness, neutral, and sadness. Upon the conclusion of the challenge, 3,589 images designated for private testing were added to the dataset, which originally contained 28,709 images for training and 3,589 images for public testing. Public test images are employed for a range of purposes in published research projects, each using the FER-2013 dataset distribution for individual training, validation, or test sets[67,68].

## 3.2 Basic emotions

**DEAP**

The main emphasis of the DEAP dataset (Database for Emotion Analysis using Physiological Signals) lies in basic emotions. The DEAP dataset [69] has been split into two sections. The first contains the ratings from 120 one-minute music video excerpts that participants reviewed online using three criteria: arousal, valence, and dominance. The participants ranged in age from 14 to 16. The second accumulation of data comprises participant ratings, physiological recordings, and face films from an experiment in which 32 participants viewed a selection of the 40 music videos mentioned above. Every participant reviewed the videos as above, and physiological and EEG signals were recorded. Oval face videos were also captured for 22 subjects.

**BP4D+**

The main focus of the BP4D+ dataset is on basic emotions. A large-scale, multimodal emotion dataset is known as BP4D+. The FERA challenge of 2017 made use of it. There are 140 participants total, aged 18–66, including 58 male and 82 female participants. Eight physiological signals are present in total: heart rate, respiration (rate and voltage), blood pressure (diastolic, systolic, mean, and raw), and electrodermal (EDA). Ten target emotions including happiness, sadness, anger, disgust, embarrassment, astonished skepticism, fear, pain, and surprise are represented in the data for each subject [70].

## 3.3 Complex Emotion Datasets

**RAF-DB**

29,672 real-world images of faces that were extracted from Flickr are included in RAF-DB. 315 talented annotators have labeled each of the RAF-DB's images, with roughly 40 independent annotators labeling each image. The single-label subset and the multi-label subset are the two distinct subsets found in RAF-DB[71].

**CEED**

The CEED (Compound Emotion Expression Database) is mainly concentrated on capturing intricate emotions. Unlike datasets that center on fundamental emotions, CEED is tailored to document and examine emotional expressions that merge multiple basic emotions or encompass more subtle and socially influenced emotional states [72]. 480 images of eight young adult actors emulating nine complicated and six basic social-sexual emotional expressions are available in the complicated Emotion Expression Database (CEED). There is some racial variety among the actors, who are both male and female. Almost 800 individuals independently scored images to confirm how the expression was perceived [73].

These datasets are crucial for advancing research in emotion recognition systems, encompassing a variety of basic and complex emotional states. It is noteworthy that in some cases, datasets primarily designed for recognizing basic emotions could also be utilized effectively for understanding and analyzing complex emotional states, thereby expanding the applicability and scope of these datasets [47, 74]. We propose eight datasets corresponding to emotion recognition systems used in recent years, including facial expression, EEG and ECG signals, which are presented in Table 1.

**Table 1.** The summarization of the mentioned datasets is listed.

| References | Dataset Name (complex/basic) | Emotions | Assessment Types Used |
|---|---|---|---|
| Zhao and Jiancheng[59] | CMED  Complex and Basic emotion | Happiness, Disgust, Fear, Anger, Sadness, Surprise, Happily surprised, Sadly surprised, Fearfully surprised, Angrily surprised, Disgustedly surprised, Happily disgusted, Sadly fearful, Sadly angry, Sadly disgusted, Fearfully angry, Fearfully disgusted, Angrily disgusted | Not specified |
| Takalkar, Madhumita A., and Min Xu [75] | CASME  Complex and Basic emotion | Contempt, Disgust, Fear, Happiness, Regression, Sadness, Surprise, Tense | Self-report ratings |
| [76] | CFEE  Complex and Basic emotion | Angry, Fearful, Disgusted, Surprised, Happy, Sad and Neutral. | Facial Action Coding System |
| Han et al [77] | FER-2013  Complex and Basic emotion | Happy, Sad, Angry, Fear, Surprise, Disgust, and Neutral | Not specified |

| [78] | DEAP<br>Basic | signal-based (EEG) | Self-report ratings, physiological recordings, EEG |
|---|---|---|---|
| Guerdelli et al [79] | BP4D+<br><br>Basic | happiness or amusement, surprise, sadness, startle or surprise, skeptical, embarrassment, fear or nervous, physical pain, angry and disgust) | Self-report ratings, physiological recordings, EEG |
| Yan et al [80] | RAF-DB<br>Complex | neutral, happy, surprise, sad, anger, disgust, fear | Not specified |
| Benda, Margaret S., and K. S. Scherf. [73] | CEED<br><br>Complex | six basic expressions (angry, disgusted, fearful, happy, sad, and surprised) and nine complex expressions (affectionate, attracted, betrayed, brokenhearted, contemptuous, desirous, flirtatious, jealous, and lovesick) | Self-report ratings |

## 4. Preprocessing in CEMRS

In this section, the challenges of preprocessing in complex emotion recognition systems (CEMRS) will be discussed. This section is divided into two subsection: a) Physical Cues Preprocessing b) Physiological Cues Preprocessing

**a) Physical Cues Preprocessing**

This section focuses on the preprocessing difficulties related to physical cues, specifically facial expressions obtained via image processing. It delves into the intricacies of preparing facial images to derive significant features that represent various emotional states. Major challenges encompass managing fluctuations in lighting conditions, facial angles, and the range of facial expressions. Furthermore, it highlights the necessity for sophisticated techniques in feature extraction to effectively capture the subtleties of intricate emotions [81].

Initial stages involve detecting and aligning faces to ensure consistent positioning and orientation of facial features in images. Following face detection, normalization methods are utilized to counteract lighting variations by adjusting brightness, contrast, and color balance for standardization. Next, noise reduction techniques like Gaussian blurring or median filtering are applied to minimize unwanted image artifacts that could disrupt emotional cue recognition. Feature extraction is then carried out to capture important facial attributes crucial for emotion recognition, including identifying facial landmarks and extracting texture descriptors from facial regions. Given that feature vectors often consist of intricate data, dimensionality reduction techniques like Principal Component Analysis (PCA) or t-Distributed Stochastic Neighbor Embedding (t-SNE) are applied to decrease computational complexity and prevent overfitting. Through these preprocessing steps, facial expression recognition systems can enhance their accuracy, resilience, and ability to generalize in emotion classification tasks. This thorough preprocessing process ensures that the recognition system is well-prepared to provide consistent and precise results across various input images, enabling applications in affective computing, human-computer interaction, and physical research [82,83].

b) Physiological Cues Preprocessing

This section will address the difficulties linked with preprocessing physiological cues, such as those derived from electroencephalography (EEG) and electrocardiography (ECG) signals through signal processing. It will encompass the preprocessing procedures needed to cleanse and filter the physiological signals, extract relevant features, and alleviate artifacts to precisely capture the underlying emotional states. Challenges in this field may involve noise reduction, artifact elimination, feature extraction from intricate physiological signals, and ensuring the dependability and precision of the processed data for emotion recognition [84].

The preprocessing workflow involves several essential steps tailored to optimize the EEG and ECG signals for effective analysis. Firstly, artifact removal is imperative to eliminate noise and unwanted interference from the signals. EEG signals, for instance, are susceptible to various artifacts like eye blinks, muscle movements, and environmental electrical noise, while ECG signals can be affected by muscle activity and movement artifacts. Techniques such as independent component analysis (ICA) and adaptive filtering are commonly used to mitigate these artifacts, ensuring that the extracted signals accurately represent the underlying neural and cardiac activity. Following artifact removal, signal segmentation is performed to divide the continuous EEG and ECG recordings into shorter, temporally meaningful epochs, often aligned with specific stimuli or events that elicit emotional responses. This segmentation allows for the analysis of emotion-related signal patterns within defined time windows. After segmentation, feature extraction is conducted to capture relevant characteristics from the EEG and ECG signals indicative of different emotional states. For EEG signals, features may include spectral power, coherence, and asymmetry in specific frequency bands, while ECG features may encompass heart rate variability (HRV), amplitude variations, and timing intervals such as RR intervals. Dimensionality reduction techniques are then used to minimize the complexity of the feature space while keeping important information, such as principal component analysis (PCA) and wavelet processing. By implementing these preprocessing steps, EEG and ECG-based emotion recognition systems can significantly improve their accuracy, robustness, and generalization capabilities, allowing for the development of more reliable and versatile emotion classification models for use in affective computing, human-computer interaction, and psychological research [85,86,87].

5. Machine Learning and deep learning Models in Emotion Recognition: Limitations and Challenges

Most studies in emotion recognition have primarily focused on utilizing machine learning, including deep learning techniques, to recognize emotions based on basic features. However, emotions are inherently complex, and traditional machine learning models, including deep learning models, often struggle to capture this complexity effectively. While these models may perform reasonably well in recognizing basic emotions, they face several limitations when it comes to understanding and categorizing complex emotional states.

5.1 Limitations of Machine/Deep Learning Models

1. **Inability to Capture Complex Emotional States:** Traditional machine learning models, including deep learning models, often rely on predefined features and lack the flexibility to capture the nuances and subtleties of complex emotional states. As a result, they may struggle to distinguish between similar emotional expressions or interpret the context-dependent nature of complex emotions [88].
2. **Dependency on Handcrafted Features:** Many traditional machine learning and deep learning approaches require handcrafted features to be extracted from the data. These features may not fully represent the multidimensional nature of complex emotions, leading to a limited ability to generalize across different emotional contexts [89,90].
3. **Limited Adaptability to Individual Differences:** Traditional machine learning and deep learning models may not be adaptable enough to account for individual differences in expressing and experiencing emotions. They often rely on generic models that do not adequately capture the variability and subjectivity of emotional responses among different individuals [91].
4. **Overfitting and Generalization Issues:** Traditional machine learning and deep learning models may be prone to overfitting when trained on limited datasets, resulting in poor generalization to unseen data. This is particularly problematic in the case of complex emotions, where the variability and diversity of emotional expressions can be significant [92].

5.2 The Role of Meta-Learning

In the face of these challenges, meta-learning offers promising approaches to enhance the capabilities of complex emotion recognition systems. Meta-learning, which focuses on learning how to learn, can provide a framework for adapting traditional machine learning and deep learning models to better handle complex emotional states. By leveraging meta-knowledge acquired from a diverse range of emotional contexts, meta-learning algorithms can enhance the adaptability and generalization capabilities of emotion recognition systems. This approach enables models to learn from previous experiences and rapidly adapt to new emotional contexts, thereby improving their performance in recognizing complex emotions [93,94].

Most studies have presented machine learning and deep learning techniques for recognizing emotion based on basic features but emotions are always complex. Few studies have worked on complex emotions based on basic features for recognition. In this scenario, another approach called meta-learning could play a vital role in these types of tasks. Fig 6. showcases our proposed taxonomy for emotion recognition systems.

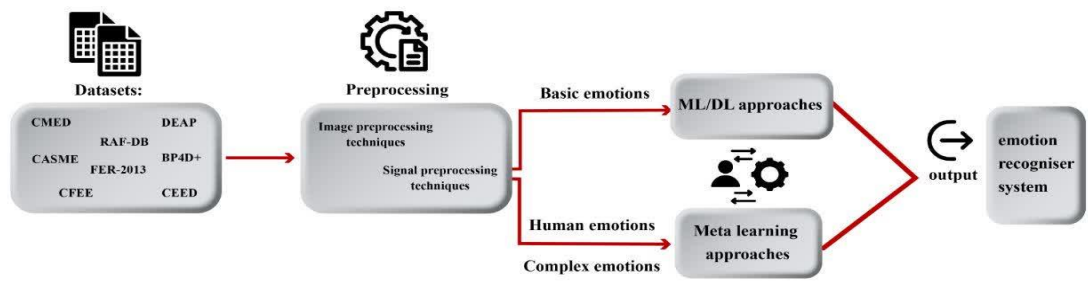

**Figure 6.** The taxonomy of emotion recognition systems.

**5.3 Complex and basic emotion recognition system (methods and technologies)**

The following sections are divided into basic and complex for introducing emotion recognizer systems based on machine learning, deep learning, and meta-learning methods.

**5.4 Machine learning, Deep learning in Basic emotion Recognition**

Machine learning approaches such as SVM, Random forest, KNN, Naive Bayes, etc in some studies utilized as feature selection or prediction tasks for emotion recognition[95]. Deep learning methods, such as recurrent neural networks (RNNs) and CNNs, have made significant strides in the field of computer vision in recent decades. These techniques based on deep learning have been used for problems related to recognition, classification, and feature extraction. By enabling "end-to-end" learning directly from input images, a CNN's primary benefit is to eliminate or greatly reduce reliance on physics-based models and/or other pre-processing approaches. Due to these factors, CNN has produced cutting-edge outcomes in several domains, such as emotion recognition tasks based on face expression, ECG, and EEG signals[96,97].

**5.5 Meta-learning in complex emotion recognition**

Meta-learning, commonly referred to as "learning to learn," is the process through which AI models acquire the ability to swiftly adjust to new environments or tasks while working with limited data. This capability is especially beneficial in the realm of complex emotion recognition, where challenges such as data scarcity and variability across different contexts frequently arise. Meta-learning comprises various methodologies: optimization-based, model-based, and metric-learning-based techniques [98].

Optimization-based techniques, including Model-Agnostic Meta-Learning (MAML), Reptile, and Almost No Inner Loop (ANIL), focus on identifying optimal initialization parameters that allow models to quickly converge on new tasks with minimal modifications. These techniques offer flexibility and adaptability across a variety of emotion recognition scenarios by effectively learning from a handful of examples [99,100].

Model-based techniques, such as recurrent and convolutional neural networks, integrate meta-learning principles directly into their architecture. These models are crafted to internally adjust to new tasks; however, they may encounter difficulties in generalizing to more complex or diverse emotional contexts due to their streamlined optimization processes.

Metric-learning-based techniques, like ProtoNet, RelationNet, and MatchingNet, depend on learning embedding functions that transform data into a space where classification can be performed using similarity metrics. These non-parametric approaches prove effective for emotion recognition as they facilitate rapid learning from sparse data by exploiting the relationships among data points [101].

In the subsequent sections, we will examine how meta-learning principles are implemented in Continual Learning, Few-shot Learning, Label Noise management, and Reinforcement Learning, all of which enhance the proficiency of models in recognizing intricate emotions:

Continual Learning guarantees that models can adapt to new emotional tasks over time without losing knowledge of previously learned tasks, which is essential in dynamic environments where emotions are subject to change.

Few-shot Learning confronts the obstacle of gaining insights from a small quantity of emotion-tagged information by empowering models to extrapolate proficiently from a tiny set of samples.

Label Noise management involves techniques for addressing noisy or inaccurately labeled emotional data, thereby improving the model's robustness and reliability in real-world emotion recognition applications.

Reinforcement Learning aids in optimizing emotion recognition policies through trial and error, allowing models to enhance their performance by engaging with their surroundings.

These methodologies underscore the importance of meta-learning in the advancement of complex emotion recognition. Through boosting the adaptability, resilience, and productivity of emotion recognition technologies, meta-learning is essential for confronting the intrinsic challenges that arise in this sector [98, 102]

**5.6 Continual learning and Few-shot learning**

Continual learning involves compiling research and methods to tackle the challenge of learning in scenarios where knowledge integration across endless streams of data needs to be considered, especially when the data distribution fluctuates over time [103,104]. In the realm of continual learning, models are crafted to acquire knowledge from new data while preserving previously gained knowledge, a task made difficult by catastrophic forgetting, a phenomenon where new learning can disrupt and overwrite previously absorbed information. To combat this issue, continual learning strategies often incorporate replay techniques, where a model is regularly trained on a combination of fresh data and selected samples of past data. These replay methods play a vital role in achieving the perfect equilibrium between stability—maintaining existing knowledge—and adaptability—effectively integrating new information. Furthermore, additional approaches like regularization techniques, which discourage alterations to crucial weights, and parameter isolation methods, which designate specific parts of the model for different tasks, are utilized to alleviate catastrophic forgetting and boost the performance of continual learning systems [105]. Fig 7. indicates the architecture of continual learning.

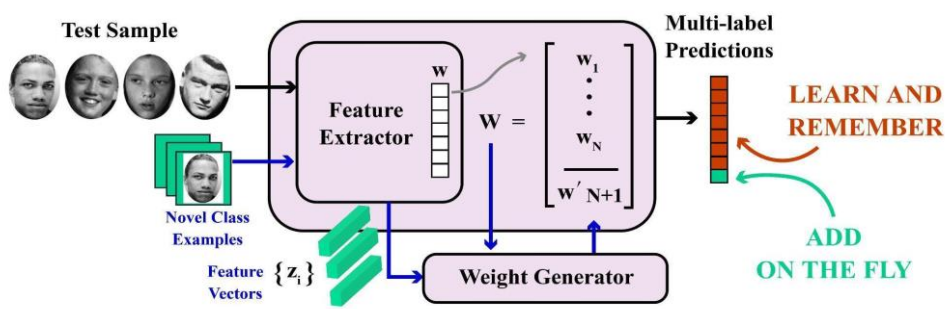

**Figure 7.** Continual learning in complex emotion recognition systems.

According to Fig 8, few-shot learning is the concept for developing an expanding algorithm from an insignificant sample set. Few-shot learning for facial emotion recognition has been established to decrease the intraclass distance and enhance the interclass distance[106].

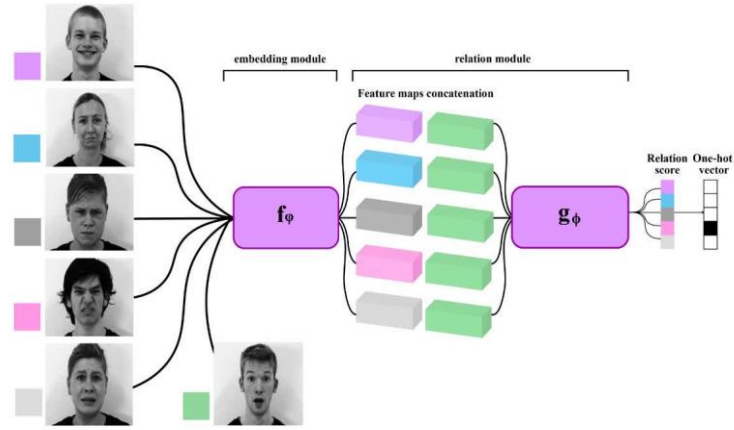

**Figure 8.** Few shot learning in complex emotion recognition systems.

### 5.7 Label noise

Label noise is a prevalent issue in real-world datasets caused by several factors, including the expense of the labeling process and the challenge of accurately classifying data. In the field of affective computing for complex emotion recognition, employing noisy labels addresses several challenges inherent in interpreting facial expressions. Concerning the subjective nature of human emotions, different annotators frequently provide diverse interpretations, leading to problems in labeling. Noisy labels help manage this ambiguity by allowing models to learn from a distribution of possible labels instead of a single, potentially erroneous one, thereby fostering more robust representations. Moreover, Fig 9 demonstrates that integrating label noise during training also enhances model robustness by leveraging techniques like label distribution learning, which improves generalization and mitigates the impact of incorrect labels. This approach is particularly beneficial for handling the complexity of real-world facial expressions, which often involve mixtures of basic emotions. Noisy labels mirror this variability and help models distinguish subtle emotional nuances. Additionally, methods like Face-Specific Label Distribution Learning (FSLDL) create augmented training samples with label distributions, broadening the range of expressions and viewpoints captured, thus enhancing the model's ability to generalize to new data. To prevent overfitting to noisy samples, techniques such as rank regularization and discriminative loss functions are employed, ensuring that the model focuses on more reliable samples and maintains overall performance. By addressing ambiguity and subjectivity, improving robustness, handling complex expressions, enhancing training with augmented data, and reducing overfitting, noisy labels significantly contribute to the advancement of complex emotion recognition systems in affective computing [107,108,109,110,111,112,113,114].

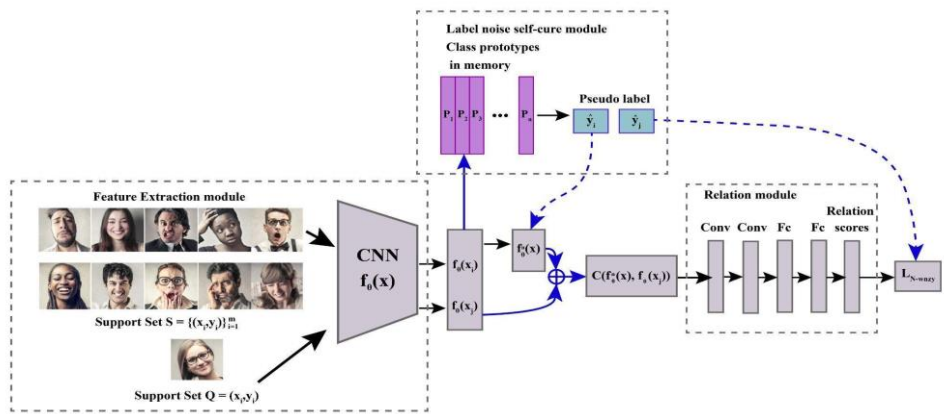

**Figure 9.** Architecture of meta-learning based on label noise.

## 5.8 Reinforcement learning

In affective computing for complex emotion recognition, employing reinforcement learning (RL) offers significant advantages by addressing key challenges. One major issue is handling unlabeled data, as complex emotions are often not explicitly labeled in datasets. RL, particularly through Deep Q-Networks (DQN), can learn from the environment via rewards and penalties, thus improving performance without the need for labeled data. Another challenge is identifying emotionally relevant intervals within data streams, such as video or physiological signals. RL-based segmentation dynamically learns to highlight these intervals, refining its strategy over time through rewards for accurate emotion recognition. It's also complex to integrate multimodal data, including facial expressions and physiological signals. RL excels in this by adaptively learning which signals are more indicative of specific emotions in varying contexts, enhancing the system's overall recognition capability.The system employs both facial expressions and physiological signals for emotion recognition. From facial expressions, it extracts confidence scores of seven basic emotions, valence-arousal (VA) levels, and ten action units (AUs). For physiological signals, it utilizes remote photoplethysmography (rPPG) to derive heart rate (HR) and heart rate variability (HRV) indices, along with EEG and ECG signals. This combination of facial and physiological data provides a comprehensive approach to recognizing complex emotions. The variability and subtlety of complex emotions often reduce recognition accuracy with traditional methods. RL optimizes segmentation and decision-making iteratively, focusing on the most informative data segments, thus improving recognition accuracy. Additionally, emotional states change rapidly, posing a challenge for static models. RL's adaptive learning allows for real-time updates in understanding and segmentation, making the system robust and effective in capturing dynamic emotional states in real-world scenarios. By addressing these challenges, RL significantly enhances the performance and reliability of complex emotion recognition systems in affective computing. Another significant advantage of using RL for complex emotion recognition is its ability to learn from sparse and unlabeled data. By focusing on key segments with significant emotional information, the RL-based segmentation module ensures that the decision module receives the most relevant data, leading to better recognition accuracy. This method is particularly effective for complex emotions, which are often subtle and not easily captured through simple observation. The RL approach allows the system to adapt and improve over time, refining its ability to detect and interpret complex emotional states. There is a step-by-step progression of emotions in interactions, which is comparable to how the action selected in reinforcement learning depends on the emotional state at that moment and the sequence of state transitions. The selected activity, representing the target's recognition results and the current emotional state, also impacts the reward function. This reinforcement learning module is essential for identifying the target emotion using characteristics of the appropriate emotion pair. By effectively managing unlabeled data, identifying key emotional intervals, integrating multimodal data, and improving recognition accuracy through iterative optimization, RL-based systems offer a comprehensive solution to the complexities of emotion recognition. The dynamic and adaptive nature of RL allows these systems to respond to rapid changes in emotional states, ensuring robust performance in real-world scenarios. The step-by-step progression and adaptation in RL mirror the evolving nature of human emotions, making it a powerful tool in affective computing. Reinforcement learning's ability to continuously improve through experience and feedback makes it particularly suited for the nuanced task of complex emotion recognition, providing a reliable and effective approach to understanding human emotions in various contexts which is highlighted in Fig 10.[115,116,117,118,119,120].

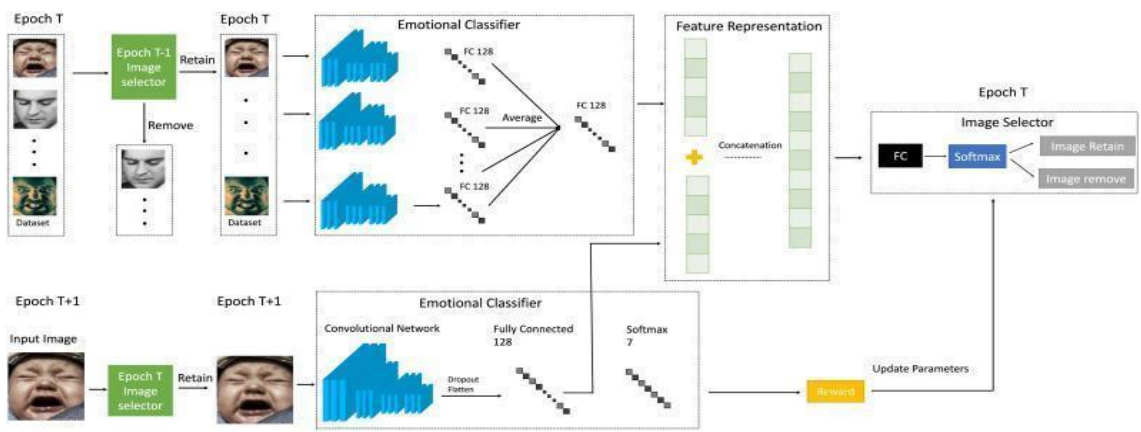

**Figure 10.** The application of reinforcement learning in complex emotion recognition systems.

## 6. A review of related works

The review of related works in basic and complex emotion recognition systems reveals a spectrum of advancements, from foundational models focusing on principal emotions. In this section, these provided studies underscore the evolution towards more accurate and context-aware systems.

### 6.1 Basic emotion recognition

In this section, we aim to review some studies related to AI approaches corresponding to the recognition of emotion by considering facial expressions, ECG, and EEG signals.

The construction of an artificial intelligence (AI) system that can recognize emotions from facial expressions. Jaiswal et al.[121] developed an AI system for emotion detection using deep learning, specifically a convolutional neural network (CNN) architecture. This model focuses on three main processes: face detection, feature extraction, and emotion classification. CNN's ability to automatically extract features and classify emotions showcases significant accuracy, achieving 98.65% on the JAFFE dataset and 70.14% on the FERC-2013 dataset. Key advantages of this model include high accuracy, the efficacy of deep learning in reducing manual feature extraction, and scalability. However, the model's limitations are its dependency on the quality and diversity of datasets, substantial computational requirements, and potential generalization issues across different populations and lighting conditions. Overall, while the CNN-based model marks a significant improvement over traditional methods, further enhancements are needed for broader application and real-world deployment.

In [122], a novel multi-modal input approach that uses color, depth, and thermal recording videos to estimate dimensional emotion states was described. Based on attention-boosted feature volumes, the proposed networks, dubbed multi-modal recurrent attention networks (MRAN), robustly distinguish facial expressions by learning spatiotemporal attention volumes. Utilizing the depth and heat sequences as guide priors, a selective focus on emotional discriminative regions on the color sequence was applied. Additionally, a brand-new multi-modal facial expression recognition benchmark called multi-modal arousal-valence facial expression recognition (MAVFER) comprised continuous arousal-valence scores matching to films of color, depth, and thermal recording. The outcomes of the experiments demonstrate that the approach was capable of producing state-of-the-art outcomes in color recording datasets for dimensional facial emotion recognition.

Zhang et al [123] studied the implementation of several deep learning models, such as long short-term memory (LSTM), convolutional neural networks (CNN), deep neural networks (DNN), and a hybrid model of CNN and LSTM (CNN-LSTM), to the subject of their research of EEG-based emotion recognition. The popular DEAP dataset was applied for evaluation. According to the experimental results, the CNN and CNN-LSTM models performed well in the categorization of EEG-based emotions, with accurate raw data extraction rates of 90.12 and 94.17%, respectively.

Bazgir et al [124] worked on the Deap dataset to build an AI system for the recognition of emotion based on electroencephalography (EEG) signals. Operating the discrete wavelet transform (DWT), EEG signals were divided into the gamma, beta, alpha, and theta frequency bands. Spectral features were then retrieved from each frequency band. To make the features mutually uncorrelated, principle component analysis (PCA) was used for the retrieved features while maintaining the same dimensionality as a transform. Emotional states were categorized by support vector machines (SVM), artificial neural networks (ANN), and k-nearest neighbors (KNN). With extracted features from ten EEG channels, the cross-validated SVM with radial basis function (RBF) kernel performs with 91.3% accuracy for arousal and 91.1% accuracy for valence in the beta frequency range.

The purpose of Sun and Lin's study [125] was to employ ECG signals to identify emotions. The data represented Four different emotions: happy, thrilling, tranquil, and tense. A finite impulse filter is then used to de-noise the raw data. To improve the accuracy of emotion recognition, The Discrete Cosine Transform (DCT) to extract characteristics from the collected data was applied. Support Vector Machine (SVM), Random Forest, and K-NN classifiers are investigated. The optimal parameters for the SVM classifier are found using the Particle Swarm Optimization (PSO) approach. The comparison of these classification techniques' findings shows that the SVM methodology recognizes emotions more accurately, which is useful in practical settings.

Nguyen et al [126] provided a new transfer learning approach utilizing PathNet to explore knowledge accumulation within a dataset and transfer insights from one emotion dataset to enhance overall performance to solve the generalization problem of developed deep learning models corresponding to a shortage of extensive emotion datasets. The proposed system by passing different series of investigations on SAVEE and eNTERFACE datasets enhances emotion recognition performance according to experimental results, outperforming recent state-of-the-art methods that employ fine-tuning or pre-trained approaches. The highest recognition accuracy that the proposed system could obtain was 93.75% on SAVEE and 87.5% on eNTERFACE.

Albraikan et al [127] worked on the E4, and MAHNOB datasets to boost the classifier's accuracy rate by utilizing peripheral physiological signals. A hybrid sensor fusion method based on a stacking model was presented which enabled the simultaneous embedding of data from several sensors and emotion models within a model that was independent of the user. As a fundamental model for classifying emotions, WMD-DTW, a weighted multidimensional DTW, and the k-nearest neighbors algorithm were employed. On top of the two base models, a high-level classifier was learned using the ensemble methods. Applying a meta-learning methodology, were able to demonstrate the ensemble method performs more effectively than any particular method. The result showed recognizing valence and arousal emotions achieved 94.0% and 93.6% employing the MAHNOB dataset.

Table 2 provides an extensive overview of the techniques, results, advantages, and disadvantages of the featured study, along with additional pertinent research. Its aim is to assist in conducting a meticulous comparison, in order to clarify the pros and cons associated with each study. In this way, it aims to enhance comprehension of the present status of research in the domain of basic emotion recognition using physical and physiological cues.

Table 2. Overview of mentioned Studies and another studies on Basic Emotion Recognition Using Physical and Physiological Cues

| **Study** | **Methodology** | **Datasets** | **Metrics/Results** | **Advantages** | **Limitations** |
| --- | --- | --- | --- | --- | --- |
| Jaiswal et al. | CNN for facial expressions | JAFFE, FERC-2013 | 70.14% (FERC-2013), 98.65% (JAFFE) | High accuracy on JAFFE | Lower accuracy on FERC-2013 |
| Sun et al. | MRAN for multi-modal input | MAVFER | State-of-the-art results | Robust to multi-modal data | Computationally intensive |
| Zhang et al. | LSTM, CNN, DNN, CNN-LSTM | DEAP | 90.12% (CNN), 94.17% (CNN-LSTM) | High classification accuracy | Potential overfitting |
| Bazgir et al. | DWT, PCA, SVM, ANN, KNN | DEAP | 91.3% (arousal), 91.1% (valence) | Effective feature extraction | Dependence on frequency bands |
| Sun and Lin | DCT, SVM, Random Forest | Four emotions (ECG signals) | Highest accuracy with SVM | High practical accuracy | Limited emotion set |
| Nguyen et al. | PathNet for transfer learning | SAVEE, eNTERFACE | 93.75% (SAVEE), 87.5% (eNTERFACE) | Improved generalization | Complexity in implementation |
| Albraikan et al. | Hybrid sensor fusion | E4, MAHNOB | 94.0% (valence), 93.6% (arousal) | Effective multi-sensor integration | High computational cost |

| Lopes et al. [128] | CNN for Facial Expression Recognition | Custom dataset | Accuracy: 94.3% | High accuracy, robust model | Limited to specific dataset |
|---|---|---|---|---|---|
| Tripathi et al. [129] | DNN, CNN for EEG Emotion Recognition | DEAP | Accuracy: 86.5% | Utilizes multiple neural networks | Limited to DEAP dataset |

**6.2 Complex emotion recognition**

In this section, an overview of selected papers focusing on meta-learning approaches for the recognition of complex emotions based on basic emotions will be presented. This review is prompted by the limited existing research on complex emotions. The selected papers are categorized into three distinct groups: Continual learning and few-shot learning will be discussed in part 1. Part 2 will cover label noises, while the investigation of reinforcement learning will be undertaken in the final part.

**Part 1) Continual learning and few-shot learning**

Angus and Nakisa [47] presented a new method based on continual learning and few-shot learning that improves and maintains its understanding of basic expression classes to recognize new compound expression classes accurately with a few training samples. Data augmentation, knowledge extraction, and a revolutionary Predictive Sorting Memory Replay to prevent catastrophic forgetting and enhance performance with fewer training samples were used. A considerable association between the activations of features in basic expressions and those in compound expressions was discovered by comparing the Grad-CAM heat maps of images of basic expressions with those of compound expressions. Continual learning outperforms non-continual learning methods in complicated face expression recognition, outperforming non-continual learning methods' state-of-the-art by 13.95%. The overall accuracy in new classes with 74.28% had demonstrated that continual learning for complex facial expression recognition played an essential role. This study is motivated by human cognition and learning patterns and it is the first to use few-shot learning for complex facial expression identification, attaining the state-of-the-art with 100% accuracy while only requiring one training sample for each expression class.

To showcasing significant improvements and practical applications in the domain of complex emotion recognition systems, Bhosale et al. [130] proposed a few-shot adaptation method from Electroencephalography (EEG) signals to address the lengthy calibration phase required by traditional Brain-Computer Interfaces (BCIs), hindering an optimal plug-and-play experience. Their model employed meta-learning to generalize well to new individuals with limited data, utilizing a few-shot learning framework trained on a small number of samples from previously unseen subjects, hence avoiding the need for retraining. Tested on the DEAP database with EEG recordings from 32 subjects watching music videos followed by emotion ratings, their method significantly improved emotion classification accuracy in terms of valence and arousal using only 20 reference samples from new subjects. Key contributions included quantifying the minimum calibration samples needed and introducing a 3-D convolutional recurrent embedding model to capture temporal relationships from spatially convolved EEG features. They explored various sampling strategies for support sets, finding that a combination of subject-dependent and subject-independent samples yielded competitive performance. In zero calibration scenarios, the model trained with subject-independent samples outperformed the supervised baseline. This system reduced the calibration burden and enhanced classification accuracy, advancing cross-subject EEG emotion recognition models and paving the way for more user-friendly and effective BCI applications.

By improving accuracy and robustness in recognizing micro-expressions for complex emotion recognition systems, a dual-branch meta-auxiliary learning method called LightmanNet for micro-expression recognition (MER) to address the challenges of limited data, subtle features, and individual differences in emotion detection was conducted by Wang et al [102]. LightmanNet utilizes a bi-level optimization process: in the first level, it learns task-specific MER knowledge through two branches. The primary branch focuses on learning MER features via primary MER tasks, while the auxiliary branch guides the model by aligning micro-expressions with macro-expressions, leveraging their spatial and temporal similarities. This dual-branch approach ensures the model learns meaningful features and avoids noise. In the second level, the model refines the task-specific knowledge, enhancing its generalization and efficiency. The method allows for quick acquisition of discriminative and generalizable MER knowledge from limited data. Extensive experiments demonstrated that LightmanNet significantly outperformed traditional, deep-learning, and meta-learning-based MER methods. The key contributions include addressing the data-level, feature-level, and decision-making-level challenges in MER, proposing a novel dual-branch meta-auxiliary learning method to improve model generalization and efficiency, and demonstrating its superior robustness and effectiveness. This work advances the field of complex emotion recognition systems by improving accuracy and robustness in recognizing micro-expressions.

**Part 2) Label Noise**

In [48], Self-cure relation networks (SCRNet), a metric-based few-shot model that is resistant to label noise and capable of classifying facial images of new classes of emotions by only a few examples from each, was introduced as a solution to the complex emotion recognition via facial expressions problem that is demonstrated in a few-shot learning problem. By generating relation scores between the query image and the sparse samples of each new class, SCRNet establishes a distance metric based on deep information abstracted by convolutional neural networks and predicts an emotion category for a query image. Six basic emotion categories such as Happiness, Surprise, Sadness, Fear, Disgust, and Anger for facial expression detection to more intricate and compound emotions had developed. Given the difficulty in obtaining large datasets and the high level of expertise required for sophisticated facial expression interpretation In order to solve the label noise issue, SCRNet uses a class prototype maintained in external memory during the Meta training phase to assign corrected labels to noisy data. The effectiveness of the proposed approach has been verified on both synthetic noise datasets as well as public datasets.

**Part 3) Reinforcement learning**

A bionic two-system architecture for recognizing complicated emotions was proposed by Wu et al [32]. The design resembles how the human brain responds to challenges and makes decisions. A quick compound sensing module is System I. System II is a slower cognitive decision-making component that interacts with data processing more. System I has two branches: one for physiological measurement, which is a practical image-only implementation, and one for facial expression feature representation, comprising fundamental emotion, action units, and valence arousal detection. In System II, a decision module with segmentation is used to verify that the chosen time includes the occurrence of the emotion and to iteratively optimize the emotion information in a particular segment via reinforcement learning. By achieving an accuracy of 94.15% for basic emotion recognition on the BP4D database with five classes and an accuracy of 68.75% for binary valence arousal classification on the DEAP, the recommended approach outperforms advanced emotion recognition tasks. The recognition accuracy on both databases exceeds 70% for a selection of complicated emotions, which is a massive improvement.

## 7. Discussion

In this study, we reviewed research papers to provide an overview of current research on meta-learning approaches, focusing on three types of unstructured data including facial expressions, EEG, and ECG signals. Considering recent developments, the domain of complex emotion systems is still in its early stages, with very few papers examining meta-learning applications in these three domains. In complex emotion recognition systems, facial expressions, along with EEG and ECG signals, play a vital role. The preprocessing steps for facial expression data are similar to signal processing techniques but are tailored for image data. Standardizing input size and removing irrelevant background information through resizing and cropping facial images help expedite processing and reduce the computational burden. Additionally, normalization techniques ensure consistent pixel intensities across images, improving model performance by mitigating potential biases. Data augmentation methods like rotation, flipping, and adding noise to facial images enhance dataset diversity and prevent overfitting, similar to signal processing counterparts. Implementing these preprocessing steps on facial expression data helps extract more meaningful features, leading to accurate and robust emotion classification. Employing facial expression data, EEG, and ECG signals in a complex emotion recognition system delivers a holistic view of an individual's emotional state. This strategy utilizes supplementary data from various modalities to improve overall performance and dependability in real-world scenarios.

### 7.1 Evaluation Metrics

The development of CERS necessitates a robust evaluation framework tailored to the unique challenges of emotion recognition, where all proposed systems must experience evaluation on standardized datasets, comparing predicted emotion ratings or labels with ground truth. In classification problems [96,131], classifiers are frequently evaluated using a confusion matrix-based technique, as shown in Fig 11 .

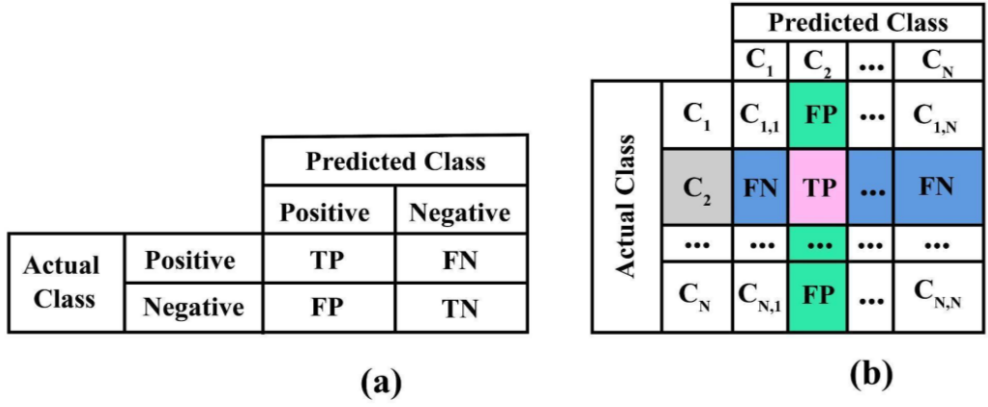

Figure 11. Confusion matrix based on CERS.

This approach provides five critical classification metrics for performance comparison; as a consequence, these criteria for the evaluation of models, including precision, recall, accuracy, specificity, and F1-score, have been constructed. Precision demonstrates the proportion of accurately detected instances of a certain emotion, emphasizing the system's ability to identify emotional states. In contrast, recall evaluates the ability of the system to detect all instances of a specific emotion in the dataset, assuring sensitivity to emotional expressions. Accuracy provides a wide overview of categorization correctness, considering true positives, true negatives, false positives, and false negatives. Specificity, represented by the True Negative (TN) rate, assesses the system's accuracy in correctly identifying instances devoid of specific emotions, crucial for distinguishing genuine emotional expressions from non-emotional states. However, in instances with unbalanced data, accuracy alone may not be sufficient, emphasizing the significance of prioritizing recall and precision for a nuanced assessment. By emphasizing recall and precision, ERS developers can better evaluate system performance, particularly in handling imbalanced datasets, leading to continuous refinement and optimization efforts in the pursuit of more accurate and reliable emotion recognition systems [132,133,134]. The Equations (1)-(5) are employed to calculate each metric, providing a comprehensive evaluation of the CERS performance.

$$\text{Accuracy} = \frac{TP + TN}{TP + FN + TN + FP} \quad (1)$$

$$\text{Precision} = \frac{TP}{TP + FP} \quad (2)$$

$$\text{Recall} = \frac{TP}{TP + FN} \quad (3)$$

$$\text{Specificity} = \frac{TN}{TN + FP} \quad (4)$$

$$\text{F1-score} = \frac{2 \cdot Precision \cdot recall}{Precision + recall} \quad (5)$$

## 8. Open Research Challenges

In the domain of complex recognition systems, there are numerous ongoing research challenges, covering a range of areas from technical enhancements to ethical considerations. One of the key challenges is the improvement of scalability and generalization capabilities in meta-learning approaches. Notably, techniques like few-shot learning and continual learning show great potential in helping models learn efficiently from limited data and adjust to changing environments. However, a major obstacle lies in scaling these approaches to handle extensive datasets and in generalizing acquired knowledge across various tasks and domains. Researchers are actively investigating new methodologies to facilitate effective knowledge transfer and adaptation, pushing the boundaries of meta-learning and broadening its relevance to real-world situations. One more critical challenge is the need to tackle label noise and shifts in data distribution, as these factors can greatly impact the performance and dependability of recognition systems. Conventional supervised learning approaches are vulnerable to the negative consequences of noisy or skewed data, leading to the requirement for the creation of effective techniques to address such obstacles. This involves exploring ways to train models that can resist label noise, creating approaches for domain adaptation, and developing algorithms that can identify and alleviate biases in data. Conquering these obstacles is essential to ensure the resilience and practicality of recognition systems in various environments and datasets. Furthermore, the interpretability, fairness, and transparency of AI models present significant research challenges, especially in situations where complex recognition systems make decisions that have a profound impact on individuals and society as a whole. It is essential to ensure that AI systems can offer understandable justifications for their decisions, reduce biases, and adhere to principles of fairness and equality to build trust and accountability. Various directions are being explored by researchers, such as explainable AI methods, fairness-aware learning algorithms, and cross-disciplinary partnerships with ethicists and social scientists, to tackle these crucial issues and encourage the responsible advancement and implementation of AI technologies. Efficiency and sustainability are crucial factors to consider in the development of intricate recognition systems. As models become more complex and computationally intensive, the focus on energy consumption and environmental impact grows. Researchers are actively seeking ways to improve model designs, create energy-efficient algorithms, and utilize hardware acceleration methods to boost the effectiveness and sustainability of AI systems. By prioritizing these initiatives, the industry can pave the path for the widespread integration of AI technologies while reducing their carbon footprint and resource usage. Finally, it is essential to carefully evaluate the broader societal impacts, such as privacy, security, and human well-being, during the creation and implementation of sophisticated recognition systems. Researchers are collaborating closely with various stakeholders to establish ethical guidelines, encourage inclusive and diverse data-gathering methods, and ensure transparent and responsible management of AI technologies. By proactively tackling these societal issues and incorporating ethical considerations into the core of AI research and advancement, the community can nurture a fairer, more accountable, and enduring future for AI technologies.

## 8.1 Conclusions and future research directions

In this study, we have delved into the effectiveness of the CERS in identifying human emotions within intricate tasks. Our research has showcased the success of CERS through different case studies, emphasizing its wide range of potential applications. This study marks a pioneering endeavor in this field, with no prior similar studies documented thus far. While previous research has focused on basic emotion recognition systems centered on facial expressions or physiological signals like EEG and ECG, our work goes beyond these methods to capture emotions within complex tasks. The case studies presented highlight the adaptability of CERS, suggesting its potential use in various sectors such as healthcare, education, and human-computer interaction. Our results indicate that even with partial implementation, CERS shows promising abilities in identifying underlying emotions linked to complex tasks. By combining data from multiple sensors, CERS can accurately deduce specific emotions, thereby improving human-computer interaction in intelligent systems. This study sets the foundation for future research aimed at improving the CERS using meta-learning techniques including incorporating advanced techniques such as few-shot learning and constant learning to allow for rapid adaptation and continuous development of emotion recognition across varied situations. Machine learning model optimization, particularly deep learning and ensemble approaches, has the potential to improve emotion recognition accuracy and robustness. Focusing on real-time emotion identification capabilities inside CERS, employing meta-learning approaches for quick adaptation to changing settings, and individualized emotion recognition models tailored to specific users' preferences are also critical. Furthermore, taking into account the ethical and societal implications of CERS deployment and defining responsible rules for its usage, including insights from meta-learning methodologies to assure fairness and transparency, is critical for the technology's ethical progress.

**References**

[1] T. Fan, S. Qiu, Z. Wang, H. Zhao, J. Jiang, Y. Wang, J. Xu, T. Sun, N. Jiang, "A new deep convolutional neural network incorporating attentional mechanisms for ECG emotion recognition," Computers in Biology and Medicine, vol. 159, pp. 106938, 2023.

[2] B. Gandhi, S. Saxena, P. Jain, "Emotion Recognition: A Review," in Microelectronics, Circuits and Systems: Select Proceedings of Micro2021, pp. 371-379, 2023.

[3] P. Ekman, "An argument for basic emotions," Cognitive & Emotion, vol. 6, pp. 169-200, 1992.

[4] P. Ekman and D. Cordaro, "What is meant by calling emotions basic," Emotion Review, vol. 3, no. 4, pp. 364-370, 2011.

[5] C. E. Izard, "Basic emotions, natural kinds, emotion schemas, and a new paradigm," Perspectives on Psychological Science, vol. 2, no. 3, pp. 260-280, 2007.

[6] K. R. Scherer, "What are emotions? And how can they be measured?," Social Science Information, vol. 44, no. 4, pp. 695-729, 2005.

[7] F. E. Oğuz, A. Alkan, T. Schöler, "Emotion detection from ECG signals with different learning algorithms and automated feature engineering," Signal, Image and Video Processing, pp. 1-9, 2023.

[8] T. Nokelainen, P. A. Airola, M. S. I. Elnaggar, "Physiological signal-based emotion recognition from wearable devices," Health Technology, 2023.

[9] B. Nakisa, M. N. Rastgoo, D. Tjondronegoro, and V. Chandran, "Evolutionary computation algorithms for feature selection of EEG-based emotion recognition using mobile sensors," Expert Systems with Applications, vol. 93, pp. 143-155, 2018.

[10] B. Nakisa, "Emotion classification using advanced machine learning techniques applied to wearable physiological signals data," Ph.D. dissertation, Queensland University of Technology, Brisbane, Australia, 2019.

[11] M. N. Rastgoo, B. Nakisa, A. Rakotonirainy, F. Maire, and V. Chandran, "Driver stress levels detection system using hyperparameter optimization," Journal of Intelligent Transportation Systems, vol. 28, no. 4, pp. 443-458, 2024.

[12] M. N. Rastgoo, B. Nakisa, F. Maire, A. Rakotonirainy, and V. Chandran, "Automatic driver stress level classification using multimodal deep learning," Expert Systems with Applications, vol. 138, 2019, Art. no. 112793.

[13] J. Kim, N. Bianchi-Berthouze, D. Patel, "Exploring User Experiences of Physical Activity Tracking Technology and the Implications for Design," Human-Computer Interaction, vol. 33, no. 3, pp. 267-304, 2018.

[14] Kranti Kamble, Joydeep Sengupta, "A comprehensive survey on emotion recognition based on electroencephalograph (EEG) signals," Multimedia Tools and Applications, vol. 82, no. 18, pp. 27269-27304, 2023.

[15] B. Majhi, N. Das, and M. Chakraborty, "Analyzing emotional responses to audio-visual stimuli through heart rate variability analysis," in 2024 IEEE International Students' Conference on Electrical, Electronics and Computer Science (SCEECS), Feb. 2024, pp. 1-6.

[16] Z. S. Chen, I. R. Galatzer-Levy, B. Bigio, C. Nasca, Y. Zhang, "Modern views of machine learning for precision psychiatry," Patterns, vol. 3, no. 11, 2022.

[17] B. M. Booth, K. Mundnich, T. Feng, A. Nadarajan, T. H. Falk, J. L. Villatte, E. Ferrara, S. Narayanan, "Multimodal human and environmental sensing for longitudinal behavioral studies in naturalistic settings: Framework for sensor selection, deployment, and management," Journal of medical Internet research, vol. 21, no. 8, p. e12832, 2019.

[18] S. Li, W. Deng, "Deep facial expression recognition: A survey," IEEE Transactions on Affective Computing, vol. 13, no. 3, pp. 1195-1215, 2020.

[19] Y. Huang, F. Chen, S. Lv, X. Wang, "Facial expression recognition: A survey," Symmetry, vol. 11, no. 10, p. 1189, 2019.

[20] N. Raut, "Facial emotion recognition using machine learning," 2018.

[21] R. W. Picard, Affective Computing, Cambridge, MA: MIT Press, 1997, p. 20.

[22] H. Wang, D. Nie, and B. L. Lu, "Emotional state classification from EEG data using machine learning approach," Neurocomputing, vol. 129, pp. 94-106, 2014.

[23] S. Koelstra, C. Muhl, M. Soleymani, J. S. Lee, A. Yazdani, T. Ebrahimi, and I. Patras, "DEAP: A database for emotion analysis using physiological signals," IEEE Transactions on Affective Computing, vol. 3, no. 1, pp. 18-31, 2012.

[24] P. Giannopoulos, I. Perikos, and I. Hatzilygeroudis, "Deep learning approaches for facial emotion recognition: A case study on FER-2013," in Advances in Hybridization of Intelligent Methods: Models, Systems and Applications, Cham: Springer, 2018, pp. 1-16.

[25] S. Pouyanfar, S. Sadiq, Y. Yan, H. Tian, Y. Tao, M. P. Reyes, M. L. Shyu, S.-C. Chen, and S. S. Iyengar, "A survey on deep learning: Algorithms, techniques, and applications," ACM Computing Surveys (CSUR), vol. 51, no. 5, pp. 1-36, 2018.

[26] A. Pramod, H. S. Naicker, and A. K. Tyagi, "Machine learning and deep learning: Open issues and future research directions for the next 10 years," in Computational analysis and deep learning for medical care: Principles, methods, and applications, 2021, pp. 463-490.

[27] D. Grandjean, D. Sander, and K. R. Scherer, "Conscious emotional experience emerges as a function of multilevel, appraisal-driven response synchronization," Consciousness and Cognition, vol. 17, no. 2, pp. 484-495, 2008.

[28] R. A. Calvo and S. D'Mello, "Affect detection: An interdisciplinary review of models, methods, and their applications," IEEE Transactions on Affective Computing, vol. 1, no. 1, pp. 18-37, Jan. 2010.

[29] M. Soleymani, S. Asghari-Esfeden, Y. Fu, and M. Pantic, "Analysis of EEG signals and facial expressions for continuous emotion detection," IEEE Transactions on Affective Computing, vol. 8, no. 3, pp. 295-308, July-Sept. 2017.

[30] J. Schmidhuber, "Deep learning in neural networks: An overview," Neural Networks, vol. 61, pp. 85-117, 2015.

[31] W. Mellouk and W. Handouzi, "Facial emotion recognition using deep learning: review and insights," Procedia Computer Science, vol. 175, pp. 689-694, 2020.

[32] Y.-C. Wu, L.-W. Chiu, C.-C. Lai, B.-F. Wu, and S. S. J. Lin, "Recognizing, Fast and Slow: Complex Emotion Recognition with Facial Expression Detection and Remote Physiological Measurement," IEEE Transactions on Affective Computing, 2023.

[33] N. Ahmed, Z. Al Aghbari, and S. Girija, "A systematic survey on multimodal emotion recognition using learning algorithms," Intelligent Systems with Applications, vol. 17, pp. 200171, 2023.

[34] S. K. Khare, V. Blanes-Vidal, E. S. Nadimi, and U. R. Acharya, "Emotion recognition and artificial intelligence: A systematic review (2014–2023) and research recommendations," Information Fusion, vol. 102019, 2023.

[35] Y. Ding, X. Tian, L. Yin, X. Chen, S. Liu, B. Yang, and W. Zheng, "Multi-scale relation network for few-shot learning based on meta-learning," in International Conference on Computer Vision Systems, Cham: Springer International Publishing, 2019, pp. 343-352.

[36] X. He, J. Sygnowski, A. Galashov, A. A. Rusu, Y. W. Teh, and R. Pascanu, "Task agnostic continual learning via meta learning," arXiv preprint arXiv:1906.05201, 2019.

[37] Z. Wang, G. Hu, and Q. Hu, "Training noise-robust deep neural networks via meta-learning," in Proceedings of the IEEE/CVF conference on computer vision and pattern recognition, 2020, pp. 4524-4533.

[38] H. Li and H. Xu, "Deep reinforcement learning for robust emotional classification in facial expression recognition," Knowledge-Based Systems, vol. 204, p. 106172, 2020.

[39] S. Gu, F. Wang, N. P. Patel, J. A. Bourgeois, and J. H. Huang, "A model for basic emotions using observations of behavior in Drosophila," Frontiers in psychology, vol. 10, p. 445286, 2019.

[40] M. H. Black, N. T. Chen, O. V. Lipp, S. Bölte, and S. Girdler, "Complex facial emotion recognition and atypical gaze patterns in autistic adults," Autism, vol. 24, no. 1, pp. 258-262, 2020.

[41] C. Zhu, P. Li, Z. Zhang, D. Liu, and W. Luo, "Characteristics of the regulation of the surprise emotion," Scientific Reports, vol. 9, no. 1, p. 7576, 2019.

[42] A. Weatherall and J. S. Robles, "How emotions are made to do things," How emotions are made in talk, vol. 321, pp. 1-24, 2021.

[43] I. Chaidi and A. Drigas, "Autism, expression, and understanding of emotions: literature review," pp. 94-111, 2020.

[44] D. Keltner and B. N. Buswell, "Embarrassment: Its distinct form and appeasement functions," Psychological Bulletin, vol. 122, no. 3, pp. 250-270, 1997.

[45] B. Mesquita and N. H. Frijda, "Cultural variations in emotions: A review," Psychological Bulletin, vol. 112, no. 2, pp. 179-204, 1992.

[46] B. Mesquita, "Emotions in collectivist and individualist contexts," Journal of Personality and Social Psychology, vol. 80, no. 1, pp. 68-74, 2001.

[47] A. Maiden and B. Nakisa, "Complex facial expression recognition using deep knowledge distillation of basic features," arXiv preprint, arXiv:2308.06197, 2023.

[48] W. G. Parrott, Emotions in Social Psychology: Essential Readings, New York, NY: Psychology Press, 2001.

[49] A. Wierzbicka, Emotions across Languages and Cultures: Diversity and Universals, Cambridge, U.K.: Cambridge University Press, 1999.

[50] A. T. Beall and J. L. Tracy, "Emotivational psychology: How distinct emotions facilitate fundamental motives," Social and Personality Psychology Compass, vol. 11, no. 2, p. e12303, 2017.

[51] A. Milone, L. Cerniglia, C. Cristofani, E. Inguaggiato, V. Levantini, G. Masi, M. Paciello, F. Simone, and P. Muratori, "Empathy in youths with conduct disorder and callous-unemotional traits," Neural plasticity, vol. 2019, no. 1, p. 9638973, 2019.

[52] R. Adolphs, "How should neuroscience study emotions? By distinguishing emotion states, concepts, and experiences," Social cognitive and affective neuroscience, vol. 12, no. 1, pp. 24-31, 2017.

[53] G. Šimić, M. Tkalčić, V. Vukić, D. Mulc, E. Španić, M. Šagud, F. E. Olucha-Bordonau, M. Vukšić, and P. R. Hof, "Understanding emotions: origins and roles of the amygdala," Biomolecules, vol. 11, no. 6, p. 823, 2021.

[54] N. B. Rothman and S. Melwani, "Feeling mixed, ambivalent, and in flux: The social functions of emotional complexity for leaders," Academy of Management Review, vol. 42, no. 2, pp. 259-282, 2017.

[55] Y. Wang, W. Song, W. Tao, A. Liotta, D. Yang, X. Li, S. Gao et al., "A systematic review on affective computing: Emotion models, databases, and recent advances," Information Fusion, vol. 83, pp. 19-52, 2022.

[56] S. PS and G. Mahalakshmi, "Emotion models: a review," International Journal of Control Theory and Applications, vol. 10, no. 8, pp. 651-657, 2017.

[57] R. Vempati and L. D. Sharma, "A systematic review on automated human emotion recognition using electroencephalogram signals and artificial intelligence," Results in Engineering, p. 101027, 2023.

[58] M. S. Chaubey and N. Pathrotkar, "Facial Recognition Ai: A Powerful Tool For Emotion Detection And Characterization," Journal of Data Acquisition and Processing, vol. 38, no. 2, pp. 1914, 2023.

[59] Y. Zhao and J. Xu, "A convolutional neural network for compound micro-expression recognition," Sensors, vol. 19, no. 24, p. 5553, 2019.

[60] W.-J. Yan, Q. Wu, Y.-J. Liu, S.-J. Wang, and X. Fu, "CASME database: A dataset of spontaneous micro-expressions collected from neutralized faces," in 10th IEEE International Conference and Workshops on Automatic Face and Gesture Recognition (FG), 2013.

[61] W.-J. Yan, X. Li, S.-J. Wang, G. Zhao, Y.-J. Liu, Y.-H. Chen, and X. Fu, "CASME II: An improved spontaneous micro-expression database and the baseline evaluation," PLoS ONE, vol. 9, no. 1, Art. no. e86041, 2014.

[62] S.-T. Liong, Y.-S. Gan, C.-J. Wong, and K. Wong, "SAMM: A spontaneous micro-facial movement dataset," IEEE Transactions on Affective Computing, vol. 9, no. 1, pp. 70-75, Jan.-Mar. 2018.

[63] A. K. Davison, C. Lansley, J. F. Cohn, H. Gunes, and B. Martinez, "SAMM: A spontaneous micro-facial movement dataset," IEEE Transactions on Affective Computing, vol. 9, no. 1, pp. 116-129, Jan.-Mar. 2018.

[64] J. Li, Y. Wang, J. See, and W. Liu, "Micro-expression recognition based on 3D flow convolutional neural network," Pattern Analysis and Applications, vol. 22, pp. 1331-1339, 2019.

[65] J. Guo, Z. Lei, J. Wan, E. Avots, N. Hajarolasvadi, B. Knyazev, A. Kuharenko et al., "Dominant and complementary emotion recognition from still images of faces," IEEE Access, vol. 6, pp. 26391-26403, 2018.

[66] C.-L. Kim and B.-G. Kim, "Few-shot learning for facial expression recognition: a comprehensive survey," Journal of Real-Time Image Processing, vol. 20, no. 3, pp. 52, 2023.

[67] G. P. Kusuma, J. Jonathan, and A. P. Lim, "Emotion recognition on fer-2013 face images using fine-tuned vgg-16," Advances in Science, Technology and Engineering Systems Journal, vol. 5, no. 6, pp. 315-322, 2020.

[68] L. Zahara, P. Musa, E. P. Wibowo, I. Karim, and S. B. Musa, "The facial emotion recognition (FER-2013) dataset for prediction system of micro-expressions face using the convolutional neural network (CNN) algorithm based Raspberry Pi," in 2020 Fifth international conference on informatics and computing (ICIC), 2020, pp. 1-9.

[69] S. Tripathi, S. Acharya, R. Sharma, S. Mittal, and S. Bhattacharya, "Using deep and convolutional neural networks for accurate emotion classification on DEAP data," in Proceedings of the AAAI Conference on Artificial Intelligence, 2017, vol. 31, no. 2, pp. 4746-4752.

[70] D. Fabiano and S. Canavan, "Emotion recognition using fused physiological signals," in 2019 8th International Conference on Affective Computing and Intelligent Interaction (ACII), 2019, pp. 42-48.

[71] A. Greco, N. Strisciuglio, M. Vento, and V. Vigilante, "Benchmarking deep networks for facial emotion recognition in the wild," Multimedia tools and applications, vol. 82, no. 8, pp. 11189-11220, 2023.

[72] Du, S., Tao, Y., & Martinez, A. M. (2014). "Compound facial expressions of emotion." Proceedings of the National Academy of Sciences, 111(15), E1454-E1462.

[73] M. S. Benda and K. S. Scherf, "The Complex Emotion Expression Database: A validated stimulus set of trained actors," PloS one, vol. 15, no. 2, p. e0228248, 2020.

[74] X. Wang, Y. Wang, and D. Zhang, "Complex Emotion Recognition via Facial Expressions with Label Noises Self-Cure Relation Networks," Computational Intelligence and Neuroscience, vol. 2023, 2023.

[75] M. A. Takalkar and M. Xu, "Image based facial micro-expression recognition using deep learning on small datasets," in 2017 international conference on digital image computing: techniques and applications (DICTA), 2017, pp. 1-7.

[76] V. Mavani, S. Raman, and K. P. Miyapuram, "Facial expression recognition using visual saliency and deep learning," in Proceedings of the IEEE international conference on computer vision workshops, 2017, pp. 2783-2788

[77] B. Han, H. Kim, G. J. Kim, and J.-I. Hwang, "Masked FER-2013: Augmented Dataset for Facial Expression Recognition," in 2023 IEEE Conference on Virtual Reality and 3D User Interfaces Abstracts and Workshops (VRW), 2023, pp. 747-748.

[78] M. Khateeb, S. M. Anwar, and M. Alnowami, "Multi-domain feature fusion for emotion classification using DEAP dataset," IEEE Access, vol. 9, pp. 12134-12142, 2021.

[79] H. Guerdelli, C. Ferrari, W. Barhoumi, H. Ghazouani, and S. Berretti, "Macro-and micro-expressions facial datasets: A survey," Sensors, vol. 22, no. 4, p. 1524, 2022.

[80] H. Yan, Y. Gu, X. Zhang, Y. Wang, Y. Ji, and F. Ren, "Mitigating label-noise for facial expression recognition in the wild," in 2022 IEEE International Conference on Multimedia and Expo (ICME), 2022, pp. 1-6.

[81] P. Schulze, A.-K. Bestgen, R. K. Lech, L. Kuchinke, and B. Suchan, "Preprocessing of emotional visual information in the human piriform cortex," Scientific Reports, vol. 7, no. 1, p. 9191, 2017.

[82] Samadiani, Najmeh, Guangyan Huang, Borui Cai, Wei Luo, Chi-Hung Chi, Yong Xiang, and Jing He. "A review on automatic facial expression recognition systems assisted by multimodal sensor data." Sensors 19, no. 8 (2019): 1863.

[83] Li, Lixiang, Xiaohui Mu, Siying Li, and Haipeng Peng. "A review of face recognition technology." IEEE access 8 (2020): 139110-139120.

[84] K. Yang, C. Wang, Y. Gu, Z. Sarsenbayeva, B. Tag, T. Dingler, G. Wadley, and J. Goncalves, "Behavioral and physiological signals-based deep multimodal approach for mobile emotion recognition," IEEE Transactions on Affective Computing, vol. 14, no. 2, pp. 1082-1097, 2021.

[85] C. R. Rashmi and C. P. Shantala, "EEG artifacts detection and removal techniques for brain computer interface applications: a systematic review," International Journal of Advanced Technology and Engineering Exploration, vol. 9, no. 88, pp. 354-383, 2022.

[86] A. Shoka, M. Dessouky, A. El-Sherbeny, and A. El-Sayed, "Literature review on EEG preprocessing, feature extraction, and classifications techniques," Menoufia J. Electron. Eng. Res, vol. 28, no. 1, pp. 292-299, 2019.

[87] C. Dora and P. K. Biswal, "Engineering approaches for ECG artefact removal from EEG: a review," International Journal of Biomedical Engineering and Technology, vol. 32, no. 4, pp. 351-383, 2020.

[88] A. Hassouneh, A. M. Mutawa, and M. Murugappan, "Development of a real-time emotion recognition system using facial expressions and EEG based on machine learning and deep neural network methods," Informatics in Medicine Unlocked, vol. 20, p. 100372, 2020.

[89] P. Tzirakis, G. Trigeorgis, M. A. Nicolaou, B. W. Schuller, and S. Zafeiriou, "End-to-end multimodal emotion recognition using deep neural networks," IEEE Journal of selected topics in signal processing, vol. 11, no. 8, pp. 1301-1309, 2017.

[90] U. Côté-Allard, E. Campbell, A. Phinyomark, F. Laviolette, B. Gosselin, and E. Scheme, "Interpreting deep learning features for myoelectric control: A comparison with handcrafted features," Frontiers in bioengineering and biotechnology, vol. 8, p. 158, 2020.

[91] J. Zhao, X. Mao, and L. Chen, "Speech emotion recognition using deep 1D & 2D CNN LSTM networks," Biomedical signal processing and control, vol. 47, pp. 312-323, 2019.

[92] E. Kanjo, E. M. Younis, and C. S. Ang, "Deep learning analysis of mobile physiological, environmental and location sensor data for emotion detection," Information Fusion, vol. 49, pp. 46-56, 2019.

[93] F. Zhou, C. Cao, T. Zhong, and J. Geng, "Learning meta-knowledge for few-shot image emotion recognition," Expert Systems with Applications, vol. 168, p. 114274, 2021.

[94] R. K. Gandhi, "Performance analysis of meta-learning and contrastive learning for speech emotion recognition."

[95] A. Sepúlveda, F. Castillo, C. Palma, and M. Rodriguez-Fernandez, "Emotion recognition from ECG signals using wavelet scattering and machine learning," Applied Sciences, vol. 11, no. 11, p. 4945, 2021.

[96] B. C. Ko, "A brief review of facial emotion recognition based on visual information," sensors, vol. 18, no. 2, p. 401, 2018.

[97] Md Rabiul Islam, Mohammad Ali Moni, Md Milon Islam, Md Rashed-Al-Mahfuz, Md Saiful Islam, Md Kamrul Hasan, Md Sabir Hossain, et al., "Emotion recognition from EEG signal focusing on deep learning and shallow learning techniques," IEEE Access, vol. 9, pp. 94601-94624, 2021.

[98] W. Wang, J. Zhang, Z. Lin, L. Cui, and X. Zhang, "Meta-learning improves emotion recognition," in Proceedings of the World Conference on Intelligent and 3-D Technologies (WCI3DT 2022), R. Kountchev, K. Nakamatsu, W. Wang, and R. Kountcheva, Eds. Singapore: Springer, 2023, vol. 323, pp. 123-135.

[99] Y. Feng, J. Chen, J. Xie, T. Zhang, H. Lv, and T. Pan, "Meta-learning as a promising approach for few-shot cross-domain fault diagnosis: Algorithms, applications, and prospects," Knowledge-Based Systems, vol. 235, p. 107646, 2022.

[100] A. Nichol and J. Schulman, "Reptile: a scalable metalearning algorithm," arXiv preprint arXiv:1803.02999, vol. 2, no. 3, p. 4, 2018.

[101] D. Nguyen, D. T. Nguyen, S. Sridharan, and others, "Meta-transfer learning for emotion recognition," Neural Computing & Applications, vol. 35, no. 13, pp. 10535–10549, 2023.

[102] Jingyao Wang, Yunhan Tian, Yuxuan Yang, Xiaoxin Chen, Changwen Zheng, and Wenwen Qiang, "Meta-Auxiliary Learning for Micro-Expression Recognition," arXiv preprint arXiv:2404.12024, 2024.

[103] Timothée Lesort, Vincenzo Lomonaco, Andrei Stoian, Davide Maltoni, David Filliat, and Natalia Díaz-Rodríguez, "Continual learning for robotics: Definition, framework, learning strategies, opportunities and challenges," Information fusion, vol. 58, pp. 52-68, 2020.

[104] Parisi, G. I., Kemker, R., Part, J. L., Kanan, C., & Wermter, S. (2019). "Continual lifelong learning with neural networks: A review." Neural Networks, 113, 54-71.

[105] Yifan Chang, Wenbo Li, Jian Peng, Bo Tang, Yu Kang, Yinjie Lei, Yuanmiao Gui, Qing Zhu, Yu Liu, and Haifeng Li, "Reviewing continual learning from the perspective of human-level intelligence," arXiv preprint arXiv:2111.11964, 2021.

[106] Songsong Tian, Lusi Li, Weijun Li, Hang Ran, Xin Ning, and Prayag Tiwari, "A survey on few-shot class-incremental learning," Neural Networks, vol. 169, pp. 307-324, 2024.

[107] Huan Yan, Yu Gu, Xiang Zhang, Yantong Wang, Yusheng Ji, and Fuji Ren, "Mitigating label-noise for facial expression recognition in the wild," in 2022 IEEE International Conference on Multimedia and Expo (ICME), pp. 1-6, IEEE, 2022.

[108] Xiang Wu, Ran He, Zhenan Sun, and Tieniu Tan, "A light CNN for deep face representation with noisy labels," IEEE Transactions on Information Forensics and Security, vol. 13, no. 11, pp. 2884-2896, 2018.

[109] Davood Karimi, Haoran Dou, Simon K. Warfield, and Ali Gholipour, "Deep learning with noisy labels: Exploring techniques and remedies in medical image analysis," Medical Image Analysis, vol. 65, p. 101759, 2020.

[110] Stanisław Saganowski, Bartosz Perz, Adam G. Polak, and Przemysław Kazienko, "Emotion recognition for everyday life using physiological signals from wearables: A systematic literature review," IEEE Transactions on Affective Computing, vol. 14, no. 3, pp. 1876-1897, 2022.

[111] Kai Wang, Xiaojiang Peng, Jianfei Yang, Shijian Lu, and Yu Qiao, "Suppressing uncertainties for large-scale facial expression recognition," in Proceedings of the IEEE/CVF Conference on Computer Vision and Pattern Recognition, pp. 6897-6906, 2020.

[112] P. Sarkar and A. Etemad, "Self-supervised ECG representation learning for emotion recognition," in IEEE Transactions on Affective Computing, vol. 13, no. 3, pp. 1541-1554, 2020.

[113] H. Song, M. Kim, D. Park, Y. Shin, and J.-G. Lee, "Learning from noisy labels with deep neural networks: A survey," in IEEE Transactions on Neural Networks and Learning Systems, 2022.

[114] J. Zeng, S. Shan, and X. Chen, "Facial expression recognition with inconsistently annotated datasets," in Proceedings of the European Conference on Computer Vision (ECCV), 2018, pp. 222-237.

[115] K. Zhang, Y. Li, J. Wang, E. Cambria, and X. Li, "Real-time video emotion recognition based on reinforcement learning and domain knowledge," in IEEE Transactions on Circuits and Systems for Video Technology, vol. 32, no. 3, pp. 1034-1047, 2021.

[116] M. Zhao and Y. Zhang, "RL-Emotion: A Deep Reinforcement Learning Framework for Multimodal Emotion Recognition in Videos," in IEEE Transactions on Affective Computing, 2021.

[117] J. Chen, H. Wu, and S. Han, "Emotion Recognition Based on Eye Movement and EEG Using Deep Reinforcement Learning," in IEEE Access, vol. 8, pp. 165899-165910, 2020.

[118] X. Liu, J. Liu, and X. Luo, "Emotion Recognition Based on Facial Micro-Expressions Using Reinforcement Learning," in Journal of Visual Communication and Image Representation, vol. 64, p. 102621, 2019.

[119] Z. Wang and T. Zhang, "Emotion Recognition Using Deep Recurrent Neural Networks with RL-Based Feature Selection," in IEEE Transactions on Cognitive and Developmental Systems, vol. 10, no. 3, pp. 668-680, 2018.

[120] Y. Zhang, Z. Zhang, and P. Li, "A Reinforcement Learning Approach to Multimodal Emotion Recognition Using Physiological Signals," in IEEE Transactions on Affective Computing, 2022.

[121] A. Jaiswal, A. K. Raju, and S. Deb, "Facial emotion detection using deep learning," in 2020 International Conference for Emerging Technology (INCET), 2020, pp. 1-5.

[122] J. Lee, S. Kim, S. Kim, and K. Sohn, "Multi-modal recurrent attention networks for facial expression recognition," in IEEE Transactions on Image Processing, vol. 29, pp. 6977-6991, 2020.

[123] Y. Zhang et al., "An Investigation of Deep Learning Models for EEG-Based Emotion Recognition," in Frontiers in Neuroscience, vol. 14, p. 622759, 2020.

[124] O. Bazgir, Z. Mohammadi, and S. A. H. Habibi, "Emotion recognition with machine learning using EEG signals," in 2018 25th National and 3rd International Iranian Conference on Biomedical Engineering (ICBME), 2018, pp. 1-5.

[125] B. Sun and Z. Lin, "Emotion recognition using machine learning and ECG signals," arXiv preprint arXiv:2203.08477, 2022.

[126] D. Nguyen, K. Nguyen, S. Sridharan, I. Abbasnejad, D. Dean, and C. Fookes, "Meta transfer learning for facial emotion recognition," in 2018 24th International Conference on Pattern Recognition (ICPR), 2018, pp. 3543-3548.

[127] A. Albraikan, D. P. Tobón, and A. El Saddik, "Toward user-independent emotion recognition using physiological signals," in IEEE Sensors Journal, vol. 19, no. 19, pp. 8402-8412, 2018.

[128] A. T. Lopes, E. D. Aguiar, A. F. De Souza, and T. Oliveira-Santos, "Facial expression recognition with convolutional neural networks: coping with few data and the training sample order," in Pattern Recognition, vol. 61, pp. 610-628, 2017.

[129] S. Tripathi, S. Acharya, R. Sharma, S. Mittal, and S. Bhattacharya, "Using deep and convolutional neural networks for accurate emotion classification on DEAP data," in Proceedings of the AAAI Conference on Artificial Intelligence, vol. 31, no. 2, pp. 4746-4752, 2017.

[130] S. Bhosale, R. Chakraborty, and S. K. Kopparapu, "Calibration Free Meta Learning Based Approach for Subject Independent EEG Emotion Recognition," in Biomedical Signal Processing and Control, vol. 72, p. 103289, 2022.

[131] X. Li, Y. Zhang, P. Tiwari, D. Song, B. Hu, M. Yang, Z. Zhao, N. Kumar, and P. Marttinen, "EEG based emotion recognition: A tutorial and review," in ACM Computing Surveys, vol. 55, no. 4, pp. 1-57, 2022.

[132] F. Javier, E. Efren, J. Miguel, O. Roberto, and G. Manuel, "Evaluation of Machine Learning Algorithms for Classification of EEG Signals," in Technologies, vol. 10, no. 4, p. 79, 2022.

[133] P. Ackermann, C. Kohlschein, J. A. Bitsch, K. Wehrle, and S. Jeschke, "EEG-based automatic emotion recognition: Feature extraction, selection and classification methods," in 2016 IEEE 18th International Conference on e-Health Networking, Applications and Services (Healthcom), 2016, pp. 1-6.

[134] D. P. Russo, K. M. Zorn, A. M. Clark, H. Zhu, and S. Ekins, "Comparing multiple machine learning algorithms and metrics for estrogen receptor binding prediction," in Molecular Pharmaceutics, vol. 15, no. 10, pp. 4361-4370, 2018.